\documentclass[aps,reprint,superscriptaddress,twocolumn,showpacs,notitlepage]{revtex4-1}
\setcitestyle{numbers,square}

\usepackage[T1]{fontenc}
\usepackage[utf8]{inputenc}
\usepackage{textcomp}
\usepackage{comment}
\usepackage{braket}
\usepackage{amsmath}
\usepackage{array,mathtools,amssymb,booktabs}
\usepackage{multirow}
\usepackage{verbatim}
\usepackage{textgreek}

\usepackage{gensymb}
\usepackage{csquotes} 

\usepackage{lipsum}

\usepackage{tikz}
\usepackage{booktabs}

\usepackage[colorlinks=true,allcolors=red,citecolor=blue]{hyperref}

\hypersetup{
  colorlinks=true,
  linkcolor	=black,
  citecolor	=blue,
  urlcolor	=blue
}

\begin{document}

\title{Exact constraints and appropriate norms in machine learned exchange-correlation functionals}

\author{Kanun Pokharel}
\email[]{kpokhare@tulane.edu}
\affiliation{Department of Physics and Engineering Physics, Tulane University,
Louisiana 70118 New Orleans, USA}

\author{James W. Furness}
\email[]{james.w.furness.1@gmail.com}  
\affiliation{Department of Physics and Engineering Physics, Tulane University, Louisiana 70118 New Orleans, USA}

\author{Yi Yao}
\affiliation{Thomas Lord Department of Mechanical Engineering and Material Science, Duke University, Durham, North Carolina 27708, USA}
\affiliation{University of North Carolina at Chapel Hill, Chapel Hill, NC 27599, USA}

\author{Volker Blum}
\affiliation{Thomas Lord Department of Mechanical Engineering and Material Science, Duke University, Durham, North Carolina 27708, USA}

\author{Tom J.P. Irons}
\affiliation{School of Chemistry, University of Nottingham, Nottingham NG7 2RD, UK}

\author{Andrew M. Teale}
\affiliation{Department of Chemistry, University of Oslo, Nottingham NG7 2RD, UK}
\affiliation{Hylleraas Centre for Quantum Molecular Sciences, Department of Chemistry, University of Oslo, P.O. Box 1033, Blindern, N-0315 Oslo, Norway} 

\author{Jianwei Sun}
\email[]{jsun@tulane.edu}
\affiliation{Department of Physics and Engineering Physics, Tulane University, Louisiana 70118 New Orleans, USA}%

\date{\today}

\begin{abstract}
Machine learning techniques have received growing attention as an alternative strategy for developing general-purpose density functional approximations, augmenting the historically successful approach of human designed functionals derived to obey mathematical constraints known for the exact exchange-correlation functional. More recently efforts have been made to reconcile the two techniques, integrating machine learning and exact-constraint satisfaction. We continue this integrated approach, designing a deep neural network that exploits the exact constraint and appropriate norm philosophy to deorbitalize the strongly constrained and appropriately normed SCAN functional. The deep neural network is trained to replicate the SCAN functional from only electron density and local derivative information, avoiding use of the orbital dependent kinetic energy density. The performance and transferability of the machine learned functional are demonstrated for molecular and periodic systems.
\end{abstract}
\maketitle

\section{Introduction}
The density functional theory (DFT) of Hohenberg, Kohn, and Sham \cite{hohenberg1964inhomogeneous,kohn1965self} allows for efficient computation of material properties by avoiding the complicated many-electron wave function in favor of the computationally convenient electron density when solving the electronic structure problem. Due to its useful accuracy and efficiency, DFT has become the most widely used computational approach for solving problems in chemistry and condensed matter physics/electronic structure. 

In the Kohn--Sham (KS) formulation of DFT, the majority of the energy is calculated exactly, leaving only a small portion of the energy, known as the exchange-correlation (XC) energy, to be approximated. There has been extensive research on improving approximations to the XC energy and the resulting functionals are roughly categorized into a hierarchy of increasing complexity and expected accuracy \cite{perdew2001jacob}. The meta-generalized gradient approximations (mGGAs), the highest category that depends only on semi-local ingredients, are becoming increasingly popular for allowing high accuracy at favorable computational cost. The mGGA functionals commonly consist of a per-particle XC energy density, $\epsilon_{\mathrm{xc}}$, built from three ingredients: the electron density, $n(\mathbf{r})$, its gradient, |$\nabla{n(\mathbf{r})}$|, and the kinetic energy density $\tau(\mathbf{r}) = \sum_i^{\mathrm{occ.}}|\nabla \varphi_i(\mathbf{r})|^2$, where $\varphi_i$ are the occupied KS orbitals. Though less common, density functionals depending on the density Laplacian $\nabla^2{n(\mathbf{r})}$ instead of (or in addition to) $\tau$ are also included at the mGGA level. The total XC energy for the system is calculated by multiplying this per-particle XC energy density by the local electron density $n(\mathbf{r})$ and integrating over all space,
\begin{align}
E_{\mathrm{xc}} &= \int d\mathbf{r}\epsilon^{\mathrm{mGGA}}_{\mathrm{xc}}(n(\mathbf{r}),|\nabla{n(\mathbf{r})}|, \tau(\mathbf{r}),\nabla^2{n(\mathbf{r})})n(\mathbf{r}), \\
 &= \int d\mathbf{r} F_\mathrm{xc}^{\mathrm{mGGA}}(n(\mathbf{r}),|\nabla{n(\mathbf{r})}|, \tau(\mathbf{r}),\nabla^2{n(\mathbf{r})})n(\mathbf{r})\epsilon_\mathrm{x}^{\mathrm{LDA}}(\mathbf{r}),
\end{align}
where $F_\mathrm{xc}^{\mathrm{mGGA}}$ is the XC enhancement factor and $\epsilon_\mathrm{x}^{\mathrm{LDA}}(\mathbf{r}) = -(3/4\pi^2)(3\pi^2n)^{1/3}$ is the exchange energy per-particle of the uniform electron gas.

The kinetic energy density, $\tau$, is commonly used in mGGAs to recognise different chemical environments through iso-orbital indicator variables \cite{Sun2013a, sun2015strongly, furness2019enhancing}, and as a component of the spherical exchange hole expansion\cite{becke1998new}. While theoretically convenient, $\tau$ introduces an implicit dependence on the KS orbitals, which  brings some complications. 1) It reduces computational efficiency by requiring additional basis function derivatives to be computed on the numerical quadrature grid, which can be more costly for Fourier transform based periodic codes. 2) It prevents the functional being used in orbital-free DFT calculations. 3) Evaluation of the XC potential for $\tau$-dependent functionals requires either optimised effective potential (OEP) techniques \cite{gorling1994exact,wu2003algebraic}, or a generalised KS scheme\cite{Seidl1996, Neumann1996,wu2003direct}. While a generalised KS treatment can be computationally convenient, the effective XC potential operator of a $\tau$ dependent mGGA is no longer a multiplicative function, $v_\mathrm{xc}(\mathbf{r})$, and is instead a non-local operator, $\hat{v}_\mathrm{xc}$.

Despite the potential advantages offered avoiding the use of orbital dependent ingredients such as $\tau$, $\nabla^2{n(\mathbf{r})}$ remains a less explored ingredient and its physical significance for the XC energy is unclear. Recently, Mejia-Rodriguez and Trickey \cite{mejia2017deorbitalization, mejia2018deorbitalized} replaced $\tau(\mathbf{r})$ with functions of $\nabla^2{n(\mathbf{r})}$ in many mGGA XC functionals to recover similar (but not identical) performance to the parent functionals. This suggests an intriguing but unclear relationship between $\tau(\mathbf{r})$ and $\nabla^2{n(\mathbf{r})}$, though an explicit relationship remains elusive despite significant effort \cite{della2016kinetic}.

Machine learning (ML) has proven to be a powerful tool for building complicated non-linear mappings for which little theoretical guidance exists. It has proved successful in building complex models across a wide variety of fields including robotics \cite{dorigo1993genetics,salichs2006maggie}, pattern recognition\cite{macleod2010time,chittka2012your}, drug design \cite{ashley2015precision,schatz2013dna,ding2014similarity}, and gaming \cite{silver2016mastering}. Within DFT research, there has been a recent practice of applying ML to construct density functionals. In 2012, Synder {\it et. al.} used a ML approximation to construct an orbital-free non-interacting kinetic energy functional $T_{\mathrm{s}}[n]$ for spinless fermion systems \cite{snyder2012finding, snyder2013orbital}. Brockherde {\it et. al.} used ML to learn the Hohenberg-Kohn (HK) map between electron density and external potential to give a mechanism that bypasses solving the KS equations \cite{brockherde2017bypassing}. Several other works have focused on the XC potential problem \cite{liu2017improving, nagai2018neural, wellendorff2012density, li2016pure}. A perspective surveying the current state of ML in computational chemistry and materials science was recently published by Westermayr, Gastegger, and Sch\"utt in Ref. \cite{westermayr2021perspective}.

The Strongly Constrained and Appropriately Normed (SCAN) functional \cite{sun2015strongly} has proven to be effective for describing a wide variety of systems \cite{sun2016accurate}, such as liquid water and ice \cite{chen2017ab}, semiconductor materials \cite{remsing2017dependence} and metal oxides \cite{gautam2018evaluating}, as well as for key properties of correlated materials like cuprates \cite{furness2018accurate,lane2018antiferromagnetic, pokharel2020sensitivity}. SCAN's success is credited to its adherence to all of the known exact-constraints applicable to a meta-GGA functional, along with the philosophy of using ``appropriate norms'' to set free parameters with minimal empiricism. These appropriate norms are provided by the systems for which a semi-local density functional approximation can be expected to be highly accurate, that is: the total energies of systems with highly localized exchange-correlation holes \cite{sun2015strongly}.

In light of this dual success of both constraint driven design and ML techniques, a question arises. Is the philosophy of exact constraints and appropriate norms compatible with ML for functional design? Growing evidence that exact constraint adherence can improve ML transferability suggests it is. Indeed, earlier works from Hollingsworth {\it et. al.} show that enforcing coordinate scaling constraints can improve machine-learned functionals \cite{hollingsworth2018can}. Nagai and collaborators \cite{nagai2022machine} recently introduced a method to analytically impose asymptotic constraints on an ML XC functional, finding generally improved accuracy. More recently, Kirkpatrick and co-workers \cite{kirkpatrick2021pushing} developed a functional DM21 (DeepMind21) that for the first time, obeys two classes of constraints on systems with fractional electrons, which are fractional charge systems and fractional spin systems\cite{kirkpatrick2021pushing}.

In our work, we explore this idea of exact constraints and appropriate norms satisfaction by training a deep artificial neural network (ANN) to reproduce the XC energy density of the SCAN functional using $\nabla^2{n(\mathbf{r})}$ instead of $\tau(\mathbf{r})$, a similar goal to the SCAN-L functional \cite{mejia2018deorbitalized}. The
de-orbitalization of SCAN stands as a convenient application for exploring the idea of constraint satisfaction in ML functionals, with SCAN-L providing an analytical benchmark for the task. 

The ML models will be trained to perform the transformation,
\begin{gather}
F_{\mathrm{xc}}^{\mathrm{SCAN}}\left(n(\mathbf{r}),|\nabla{n(\mathbf{r})}|,\tau(\mathbf{r})\right)\nonumber\\
\Big\downarrow\label{eq:mapping}\\
F_{\mathrm{xc}}^{\mathrm{SCAN-ML}}\left(n(\mathbf{r}),|\nabla{n(\mathbf{r})}|, \nabla^2{n(\mathbf{r})}\right)\nonumber
\end{gather}
We approach this mapping of orbital free ingredients onto the SCAN XC energy density using two different ML models adhering to different numbers of exact conditions. One model is a single totally-connected ANN trained for Eq. \ref{eq:mapping} directly, termed the ``combined model''. The other model is built as two complementary exchange and correlation-like ANNs designed to obey exact spin-scaling constraints, termed the \enquote{spin-scaled model}. We also impose the general Lieb--Oxford bound \cite{lieb1981improved} on these models to create models that also satisfy this constraint.

\section*{Exact Constraints}
While the exact XC functional remains unknown, it is known to obey many mathematical conditions, commonly called the ``exact constraints'' of XC functionals. Currently, 17 exact constraints  \cite{sun2015strongly}, are known to apply at the semi-local functional level. These can be broken down as conditions for the exchange energy: (1) negativity, (2) spin-scaling \cite{oliver1979spin}, (3) uniform density scaling \cite{levy1985hellmann}, (4) the slowly-varying density gradient expansion (to fourth order) \cite{svendsen1996gradient}, (5) non-uniform density scaling \cite{pollack2000evaluating,perdew1992accurate}, and (6) a tight bound for two-electron densities \cite{perdew2014gedanken,lieb1981improved}. For correlation: (7) non-positivity, (8) the slowly-varying density gradient expansion (to second order) \cite{perdew2018erratum}, (9) uniform density scaling to the high-density limit \cite{levy1985hellmann}, (10) uniform density scaling to the low-density limit \cite{levy1985hellmann}, (11) zero correlation energy for any one-electron spin-polarized density, and (12) nonuniform density scaling \cite{pollack2000evaluating,perdew1992accurate}. Finally, there are constraints known for the exchange and correlation together: (13) size extensivity, (14) the general Lieb--Oxford bound \cite{Perdew1996,perdew1991electronic,lieb1981improved}, (15) weak dependence upon relative spin polarization in the low-density limit \cite{seidl2000simulation,tao2003climbing}, (16) static linear response of the uniform electron gas \cite{tao2008nonempirical}, and (17) the Lieb--Oxford bound for two-electron densities \cite{lieb1981improved}. 

Here, we consider a subset of the 17 constraints that are easy to enforce in an ML model. The first of these is the behavior of the exchange energy under constraint (3): uniform density scaling,
\begin{equation}
n_{\gamma}(\mathbf{r}) = \gamma^{3}n(\gamma\mathbf{r}),  \label{eq:den_scale}
\end{equation}
where $\gamma$ is a positive real number. 
The exact exchange energy is known to scale as, \begin{equation}
E_{\mathrm{x}}[n_{\gamma}] = \gamma E_{\mathrm{x}}[n],    \label{eq:ex_scale}
\end{equation}
under this transformation. The effect of this condition on kernel ridge regression models was investigated by Hollingsworth, Baker, and Burke in Ref. \citenum{hollingsworth2018can} for Hooke's atom model systems, concluding that its inclusion improved ML functional performance.

Constraint (2), the spin-scaling relation for exchange energy,
\begin{equation}
E_{\mathrm{x}}[n_{\uparrow}, n_{\downarrow}] = \frac{E_{\mathrm{x}}[2n_{\uparrow}]+E_{\mathrm{x}}[2n_{\downarrow}]}{2},
\end{equation}
is simple to enforce for ML exchange models by requiring separate exchange and correlation models, and that the same exchange model handle each spin channel independently.

Constraint (14), the general Lieb--Oxford bound on the XC enhancement factor, states that,
\begin{equation}
0 \leq F_{\mathrm{xc}} (\mathbf{r}) \leq 2.215.
\end{equation}
These bounds can be enforced on ML models by including a post-processing step that maps the ML model output, denoted $\mathrm{ANN}_{\mathrm{xc}}(\mathbf{r})$, to the desired domain, e.g.
\begin{equation}
F_{\mathrm{xc}}^{\mathrm{ML}}(\mathbf{r}) = \frac{2.215}{1+\mathrm{ANN}_{\mathrm{xc}}(\mathbf{r})^{2}}.\label{eq:lb}
\end{equation}
A similar approach can be applied to impose constraint (6), the tight bound for the exchange enhancement factor $F_{\mathrm{x}} (\mathbf{r}) \in [0,1.174]$, if the exchange and correlation models are separated. Conveniently, such post-processing also enforces constraints (1) and (7), non-positivity, by constraining $F_\mathrm{xc} \geq 0$. 
It appears more challenging to enforce the exact constraints outside this subset in ML models. For example, while enforcing the second (and fourth) order gradient expansions for correlation (and exchange) is relatively straightforwards in analytical functionals, the ML design contains thousands of parameters which cannot be fully controlled. Thus, it is non-trivial to enforce such gradient expansion constraints on the model \emph{a priori}. Despite this, the nature of supervised training against methods that obey such constraints (such as SCAN) will result in the trained model effectively learning the constraints to some degree. However, without the rigorous enforcement described above it is unclear how well such adherence will transfer out of the training domain.

\section*{Input domain}
Identifying the input domain is a critical part of ML model design as a model's performance can be strongly dependent upon the nature of its inputs. Since our central interest is to replace the kinetic energy density $\tau(\mathbf{r})$ dependence we will only consider orbital free ingredients. Four density inputs were initially identified,
\begin{align}
r_{\mathrm{s}} = \left(\frac{3}{4{\pi}n}\right)^{1/3} &- \text{Wigner-Seitz radius} \label{eq:rs}\\
s = \frac{|\nabla n|}{2(3\pi^2)^{1/3}n^{4/3}} &- \text{Reduced density gradient} \\
\zeta = \frac{n_{\uparrow}-n_{\downarrow}}{n_{\uparrow}+n_{\downarrow}} &- \text{Spin polarization} \\
q = \frac{\nabla^2n}{4(3\pi^2)^{2/3}n^{5/3}} &-  \text{Reduced density laplacian} \label{eq:q}
\end{align}

The Weigner--Seitz radius is the radius of a sphere which on average contains for a uniform density n. The reduced density gradient introduces inhomogeneity which measures how fast and how much the density varies on the scale of the local Fermi wavelength $2\pi/k_\mathrm{F}$ where $k_\mathrm{F} = (3\pi^2n)^{1/3}$ is the Fermi wavevector. The reduced density Laplacian also measures the density inhomogeneity and can distinguish bonds, in contrast to reduced gradient which vanishes at the middle of the bond \cite{della2016kinetic}. The above dimensionless ingredients are preferred for XC functionals rather than using the density variables directly as the correct uniform coordinate density-scaling behaviour can be satisfied with them in conventional XC functionals \cite{levy1985hellmann}. We note however that the non-linear manipulations of these variables made by the ML models likely break the formal scaling limits in practice. Particularly, we supply density information to the exchange model through $r_s$, thereby violating the exchange coordinate-scaling condition.
 
Further exploration revealed that including additional ingredients directly from the SCAN exchange and correlation functionals (see supplemental material of Ref. \citenum{sun2015strongly}) could improve model performance:
\begin{align}
    \epsilon_{\mathrm{c}}^0(r_{\mathrm{s}},s, \zeta)&- \text{Single orbital correlation}\label{eq:ec0}\\ 
    \epsilon_{\mathrm{c}}^1(r_{\mathrm{s}},s, \zeta)&- \text{Slowly varying correlation} \\
    g_{\mathrm{x}}(s)&- \text{Exchange inhomogeneity} \\
    h_{\mathrm{0}} = 1.174&- \text{Single orbital exchange} \\
    h_{\mathrm{1}} = {1.065}&-\frac{0.065}{\left(1+\frac{10s^2/81}{0.065}\right)}- \text{X $2^{\mathrm{nd}}$ order gradient expansion} \label{eq:h1x}
\end{align}

These additional inputs are combinations of the original density ingredients and do not provide any new information directly. Their inclusion makes learning more efficient however, as it reduces the manipulations that the network must learn. This limits dependence on the network to only learning $\tau$ dependent aspects, rather than requiring it to learn every detail of the SCAN functional.

The possible range of the input parameters is very different from the desired range of the model outputs: $0 \leq F_\mathrm{xc}^{\mathrm{ML}} \leq 2.215$. For example, the domain of $r_s$ and $s$ is $[0, +\infty)$ while $q$ is $(-\infty, \infty)$. Such a mismatch in the magnitude of input and output is known to be challenging for ML models. To correct for this we pre-processed the unbounded inputs using the hyperbolic tangent function, $\tanh(x)$ \cite{kalman1992tanh}, to smoothly map the unbounded quantities to $(-1, +1)$. With pre-processing, the inputs are defined as,
\begin{align}
\tilde{r}_{\mathrm{s}} &= \mathrm{tanh}(r_{\mathrm{s}}),\label{eq:rs_pp} \\
\tilde{s} &= \mathrm{tanh}(s), \label{eq:s_pp} \\
\tilde{q} &= \mathrm{tanh}(q).
\end{align}
Finally, we pre-process $\zeta$ as,
\begin{equation}
    \tilde{\zeta} = \frac{1}{2}\left[(1+\zeta)^{4/3}+(1-\zeta)^{4/3}\right], \label{eq:zeta_pp}
\end{equation}
to ensure the model is a symmetric function of spin polarisation \cite{Oliver1979}. The inputs of Eqs. \ref{eq:ec0}-\ref{eq:h1x} are unprocessed as their ranges are already correctly bounded. Note that the pre-processed variables (Eqs. \ref{eq:rs_pp}-\ref{eq:zeta_pp}) are only supplied to the network and are not used to generate the additional inputs of Eqs. \ref{eq:ec0}-\ref{eq:h1x}.

Having identified the input domain, a training data set consisting of 20 atoms was generated using accurate spherical Hartree--Fock Slater-type orbitals \cite{Clementi1974, Koga1999, Furness2021a} containing open and closed shell atoms (He, Li, Be, B, C, N, O, F, Ne, Na, P, Cl, Ar, K, Cr, Cu, Cu$^+$, As, Kr, Xe) with $s$, $p$, and $d$ valence shells.  Spherical Hartree--Fock Slater-type orbitals were chosen to match the data used to fit parameters in SCAN, though we expand the set of atoms significantly beyond the rare gas atoms used to fit SCAN. The energetically important region of the atomic density is typically between 0 and 4 $a_0$ and the density of each atom was therefore sampled in shells of decreasing sample density for models, with 2500 radial points uniformly sampling in $r < 1$ bohr (core), 1500 in $1 \leq r < 4$ bohr, and 500 in the tail region $4 \leq r < 10$ bohr.

This atomic training data was augmented with densities from the ``appropriate norm'' systems used in SCAN's construction. The first norm is the one electron hydrogen atom, which is used to ensure that SCAN is one-electron self interaction free (constraint 11). The second and third norms are the jellium surface densities for ${r}_{\mathrm{s}}$ =  2, 3, 4 and 6 \cite{Almeida2002, Wood2007}, and the converged SCAN orbitals of the compressed argon dimer with nuclear separations of 1.6, 1.8, and 2.0 \AA, which were used to fix SCAN's interpolation function parameters \cite{sun2015strongly}. We restrict the training data set to only these appropriate norms and avoid including chemically bonded systems. While increasing the domain of the training data is expected to give the resulting ML models higher accuracy for a wider range of problems, our intent here is to observe how exact constraint satisfaction can transfer knowledge from minimal training data onto diverse problems.

\begin{figure*}[t]
\includegraphics[width=1\linewidth]{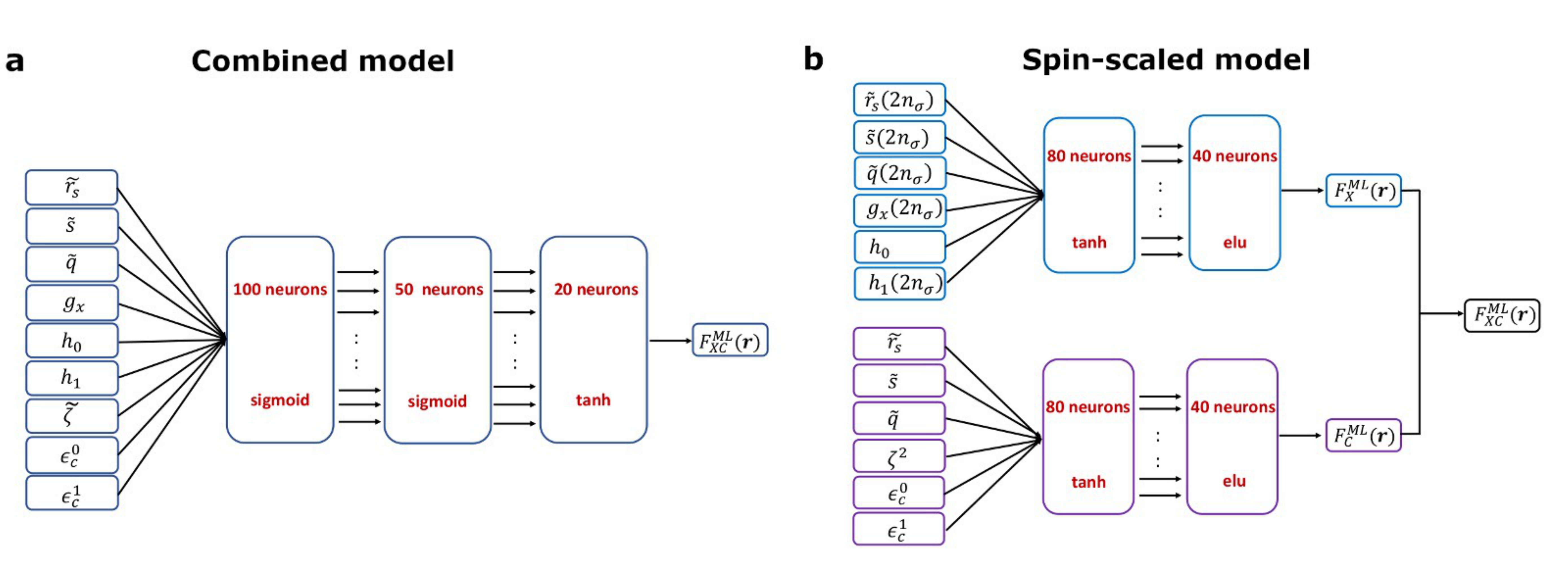} 
\caption{ML model architecture and workflow for $(\textbf{a})$ combined model $(\textbf{b})$ spin-scaled model. For the combined model, total density $(n_\uparrow + n_\downarrow)$ is given as the input. In the spin-scaled model, the upper architecture is for the exchange model learning while the lower is the correlation-like model. Spin-scaling is satisfied in the exchange model represented by $(2n_\sigma)$ where $\sigma = \uparrow \downarrow$ spin channels.}
\label{fig:ml}
\end{figure*}

\section*{Neural Network architecture and Training}
In this work, all the networks were built on the basis of the ML framework Tensorflow \cite{abadi2016tensorflow}. For the models, the non-linearity in the mapping is acquired by using sigmoid\cite{han1995influence}, tanh and exponential linear unit (elu) \cite{clevert2015fast}  activation functions, chosen as commonly used continuously differentiable activation functions. Ensuring smooth activation functions was found to be essential for obtaining reasonable XC potentials, as discussed below. The data set was randomly divided into a training (80\%) set and a validation (20\%) set, using train-test split feature of sci-kit learn \cite{scikit-learn}. Hyper-parameter searching identified a 3 layered model with sigmoid to sigmoid to tanh activation functions as preferable for the combined model while a 2 layered model with tanh to elu activation functions was preferred for the spin-scaled model. The network weights and biases were optimized by stochastic gradient descent with the Adam optimizer \cite{kingma2014adam} using a learning rate of 0.05, applying a gradient step after each sample in the training data set. The optimized model was chosen as that which minimizes the error for the validation set, generally found after one complete pass of the training data.

\subsection*{Combined Model}

Figure \ref{fig:ml} a) presents a schematic for the simple neural network(NN) architecture, termed the ``combined model''. As the name suggests, the combined model receives inputs constructed from  total density ($\mathrm{n}_{\uparrow} + \mathrm{n}_{\downarrow}$) as the features, and targets SCAN's exchange-correlation enhancement factor. Different numbers of hidden layers and neuron counts were tested, with a three layered model with 100, 50 and 20 neurons in the respective layers found to perform best.

For the combined model, the loss function to be optimized in the learning process is defined as,
\begin{equation}
\mathcal{L}_{\mathrm{combined}}= \frac{1}{N}\sum_i^N \left(F_{\mathrm{xc}}^{\mathrm{ML}}-F_{\mathrm{xc}}^{\mathrm{SCAN}}\right)^{2},
\end{equation}
where $N$ is the number of training data points, thus minimizing the mean square difference between SCAN XC enhancement and the learned XC enhancement.The Lieb-Oxford bound for the combined model is introduced as a post-processing mechanism following the explanation in Eq. \ref{eq:lb}.

\subsection*{Spin-Scaled Model}

The spin-scaled model follows a more complex architecture that allows it to obey the spin-scaling exact constraint by treating exchange and correlation separately, as discussed above. The overall architecture for the spin-scaled model is presented in Figure \ref{fig:ml} b). The spin-scaled model is comprised of two separate networks, one for exchange and one for correlation. These networks are trained separately and later combined to form the complete model. Training is therefore carried out as a two step process.

This separation by spin channel reduces the input domain for the exchange network to six features suitable for exchange, \{$\tilde{r}_{\mathrm{s}\sigma}, \tilde{s}_\sigma, \tilde{q}_\sigma, g_{\mathrm{x}\sigma}, h_{\mathrm{0}\sigma}, h_{\mathrm{1}\sigma}$\}, separately generated for each spin $\sigma$. The exchange network is trained first to minimize the mean square difference with the SCAN exchange enhancement defined as,
\begin{equation}
\mathcal{L}_{\mathrm{exchange}}= \frac{1}{N}\sum_i^N\left(F_{\mathrm{x}}^{\mathrm{ML}}-F_{\mathrm{x}}^{\mathrm{SCAN}}\right)^{2},
\end{equation}
with the spin-scaled exchange enhancement,
\begin{equation}
F_{\mathrm{x}}^{\mathrm{ML}} = \frac{F_{\mathrm{x}}(2n_{\uparrow})e_{\mathrm{x}}^{\mathrm{LDA}}(2n_{\uparrow}) + F_{\mathrm{x}}(2n_{\downarrow})e_{\mathrm{x}}^{\mathrm{LDA}}(2n_{\downarrow})}{2e_{\mathrm{x}}^{\mathrm{LDA}}(n_{\uparrow}+n_{\downarrow})}.
\end{equation}
As the exchange energy must be invariant to permutation of spin labels, the same exchange network is used for both spin channels and should be trained on both spin channels of the training data. The exchange network has two layers with 80 and 40 neurons at the first and second layers respectively and the activation functions are tanh and exponential linear unit (elu) \cite{clevert2015fast}.

The correlation energy is not subject to the same spin scaling constraint and is handled by a separate model. This second model takes a reduced set of the total density ($\mathrm{n}_{\uparrow} + \mathrm{n}_{\downarrow}$) input variables suitable for correlation: \{$\tilde{r}_{\mathrm{s}}, \tilde{s}, \tilde{q}, \zeta^{2}, \epsilon_{\mathrm{c}}^{\mathrm{0}}, \epsilon_{\mathrm{c}}^{\mathrm{1}}$\}. This second network has the same hyper-parameters as the exchange network.

\begin{table}[!ht]
    \centering
    \caption{Deviation of the spin-scaled exchange network total exchange energy (hartrees) from the SCAN exchange functional against which it was trained. Calculated for the set of spherical atom densities and the molecules of the G3 test set. Mean error (ME) and Mean absolute error (MAE) are presented, full data is given in supplemental materials. \label{tab:xonly}}
    \begin{tabular}{ccrr}
    \hline
        ~ & ~ & No LO Bound & LO Bound \\ \hline
        \multirow{2}{*}{Atoms} & ME &-0.147&-0.541\\
         & MAE&0.153&0.543\\ \hline
        \multirow{2}{*}{G3} & ME & -0.209 & -0.448 \\
         & MAE & 0.209 & 0.448 \\ \hline
    \end{tabular}
\end{table}

The loss function for the second network is,
\begin{equation}
\mathcal{L}_{\mathrm{correlation}}= \frac{1}{N}\sum_i^N\left[F_{\mathrm{c}}^{\mathrm{ML}} - (F_{\mathrm{xc}}^{\mathrm{SCAN}} - F_{\mathrm{x}}^{\mathrm{ML}})\right]^{2},\label{eq:ML_c_L}
\end{equation}
where $F_{\mathrm{x}}^{\mathrm{ML}}$ is the output of the previously trained exchange network. This second network is therefore not a true model of SCAN correlation as the loss function of Eq. \ref{eq:ML_c_L} drives it to compensate for deficiencies in the exchange network, though correlation effects will dominate if the exchange network is accurate. Table \ref{tab:xonly} shows the deviation of the bounded and un-bounded exchange networks in total exchange energy from SCAN for the spherical atoms which were part of the training and test sets, and the molecules of the G3 set which were not in training set. From this we see that deviation is typically $<1\%$ for both networks, though the bounded model deviates more significantly than the unbounded, indicating that good exchange models have been learned.

Finally, the total enhancement factor is obtained by summing the exchange network and second network enhancement factors as,
\begin{equation}
F_{\mathrm{xc}}^{\mathrm{spin-scaled}} = F_{\mathrm{x}}^{\mathrm{ML}} +  F_{c}^{\mathrm{ML}}.
\end{equation}

The Lieb-Oxford bounds for the spin-scaled model are introduced separately for exchange and the correlation parts as they are trained separately. For the exchange part, we follow similar mechanism as explained in Eq. \ref{eq:lb} where the bound is introduced to the exchange enhancement factor as,
\begin{equation}
F_{\mathrm{x}}^{\mathrm{ML-bound}}(\mathbf{r}) = \frac{1.174}{1+\mathrm{ANN}_{\mathrm{x}}(\mathbf{r})^{2}}. \label{eq:lo_x}
\end{equation}
Here we choose a tight bound of 1.174 for the exchange.

For correlation, we introduce the bound by modifying the loss function to include an additional penalty term,
\begin{multline}
\mathcal{L}_{\mathrm{correlation}}^\mathrm{{bound}}= \frac{1}{N}\sum_i^N\{\left[F_{\mathrm{c}}^{\mathrm{ML}} - (F_{\mathrm{xc}}^{\mathrm{SCAN}} - F_{\mathrm{x}}^{\mathrm{ML}})\right]^{2}\\ +\mu\times\mathrm{relu}[F_{\mathrm{xc}}^{\mathrm{ML}} - 2.215]\},\label{eq:ML_c_L_bound}
\end{multline}
where $\mu$ value is chosen to be 20, and relu is the rectified linear unit\cite{nair2010rectified}. Here we see that if the total XC enhancement factor is smaller than the tight bound 2.215 the penalty term is zero whereas any value of total XC enhancement factor greater than 2.215 will incur penalty. This total loss function is minimized to satisfy the Lieb--Oxford bound introduced in the ML model.

\section*{Results and discussion}
\subsection*{Atomic Performance}

\begin{figure}[h!] 
\includegraphics[width=0.99\linewidth]{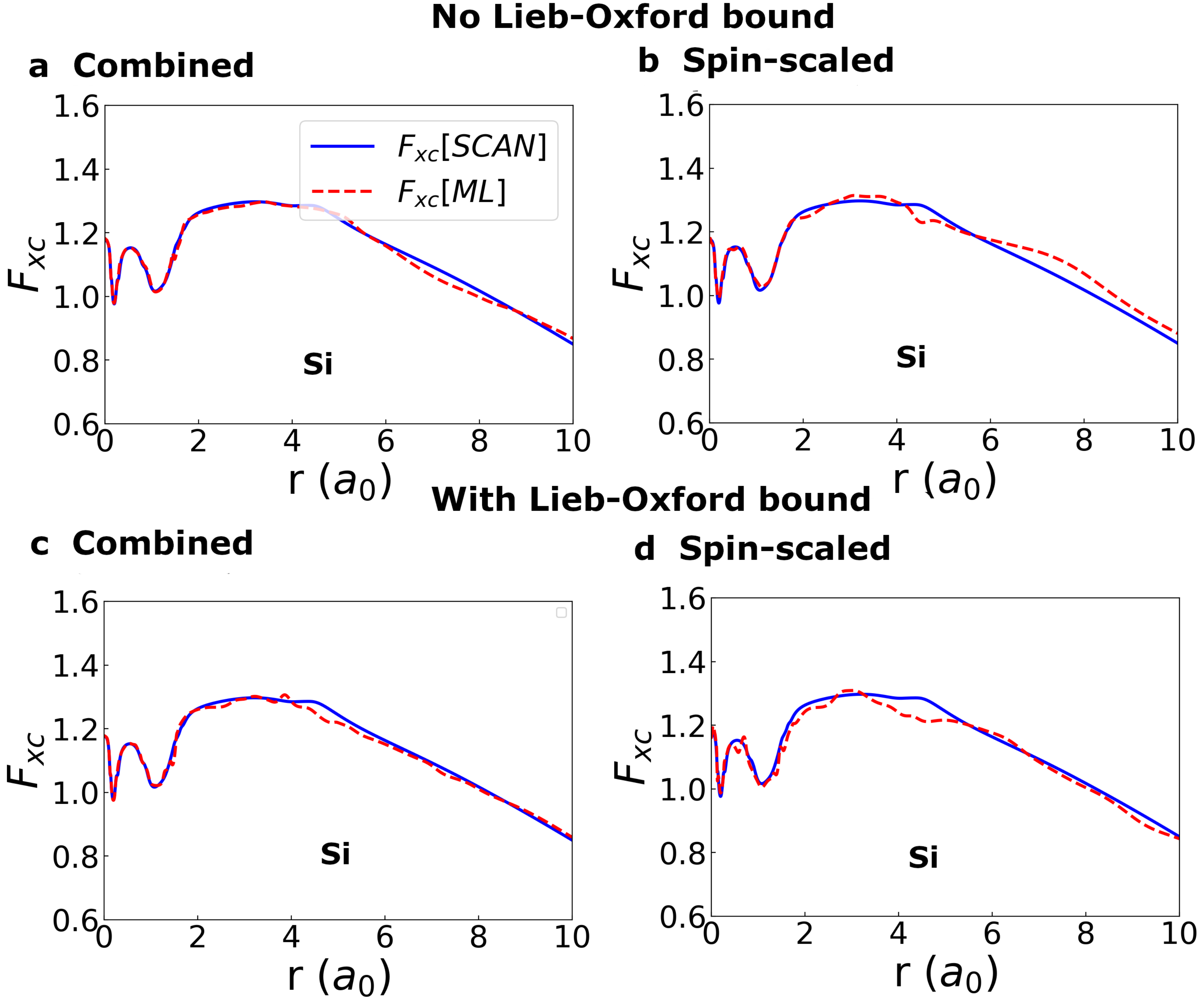} 
\caption{XC enhancement plots for the test silicon atom which was not included in the training set for $(\textbf{a})$ combined model and $(\textbf{b})$ spin-scaled model without Lieb--Oxford bound while $(\textbf{c})$ and $(\textbf{d})$ represent same as $(\textbf{a})$ and $(\textbf{b})$ with Lieb-Oxford bound. All models are compared against SCAN functional. Density was obtained from accurate spherical HF orbitals \cite{Clementi1974, Koga1999, Furness2021a}.}
\label{fig:atomic}
\end{figure}

Figure \ref{fig:atomic} shows the SCAN and ML-model XC enhancement for the silicon atom, which was not part of the training set. Figures a) and b) show combined and spin-scaled models respectively without the Lieb--Oxford bound, while c), d) include the Lieb--Oxford bound constraint. All the ML models show good agreement with the SCAN's XC enhancement factor for this system. The combined model without the Lieb--Oxford bound shows slightly less variation in the energetically important region between r = 0 and r = 4.

During training we did not target the total XC energy directly, in favor of learning the XC energy density of an existing mGGA instead. This switch towards learning the XC energy density, a local property, has three benefits. Most importantly it reduces the complexity of the training by avoiding summing the derivatives of many training points in a numerical integration batch. Secondly, far more training data is available for a given functional’s local energy density than for total XC energies, as every point in the all-space integration of any system can now be considered a training point. We note that all training points were weighted equally. As some points are more energetically important than others this will implicitly bias training towards energetically important regions, such as core densities, and away from less important regions, such as asymptotic densities. This could be balanced by dividing the training weight of each point by the uniform electron gas exchange energy for the density at the point, but we did not explore this here. Thirdly, this avoids the introduction of a gauge freedom in which many different energy density functions can integrate to the same total energy, which could result in learning a model that gives reasonable total energy, but poor local accuracy to the SCAN energy density.  Despite this training against local XC density it will be shown that the models are successful in recovering the total XC energy for the training atom sets, predicting the global property from local training. 

\subsection*{Molecular Test Sets}

For the models trained against data from atomic systems, a real challenge is to generalize to problems outside the training domain. We examine this by looking at model accuracy for the open- and closed-shell molecules of the G3 test set \cite{Curtiss2000}. The input ingredients (density, gradient and laplacian) and the SCAN $F_\mathrm{xc}^\mathrm{SCAN}$ target for all molecular calculations were generated from self-consistent SCAN orbitals in the 6-311++G(3df,3pd) basis set \cite{clark1983efficient, frisch1984self}. All molecular calculations were carried using the QUantum Electronic Structure Techniques (QUEST) program \cite{QUEST}.

\begin{table}[h!]
\caption{Mean absolute error (MAE) in kcal mol$^{-1}$ for G3 set of 226 molecular atomization energies \cite{Curtiss2000}. All SCAN G3 calculations were performed fully self-consistently with the 6-311++G(3df,3pd) basis set \cite{clark1983efficient, frisch1984self} in the QUEST program\cite{QUEST}. The ML calculations were performed non self-consistently from SCAN orbitals. The ML models with Lieb–Oxford bound are denoted by “LO” within table. Errors are given as relative to the total energy calculated using the parent SCAN functional.}
\resizebox{\columnwidth}{!}{
\begin{tabular}{c   c c c}
\hline
\multicolumn{4}{c}{\textbf{SCAN Comparison}}\\
ML models &closed shell&open shell&(G3)\\
&(MAE)&(MAE)&(MAE)\\
\hline
Combined&7.49 &7.89&7.69\\
Combined-LO&11.88 &9.26&10.58\\
Spin--scaled&8.03&4.85&6.44\\
Spin-scaled-LO&7.66 &5.66&6.66\\
\hline
 \end{tabular}
 \label{table:scan}
 }
\end{table}

\begin{table}[h!]
\caption{Mean absolute error (MAE) in kcal mol$^{-1}$ for G3 set of 226 molecular atomization energies \cite{Curtiss2000}. All SCAN G3 calculations were performed fully self-consistently with the 6-311++G(3df,3pd) basis set \cite{clark1983efficient, frisch1984self} in the QUEST program\cite{QUEST}. The ML calculations were performed non self-consistently from SCAN orbitals. The ML models with Lieb–Oxford bound are denoted by “LO” within table. Errors are given relative to standard reference values for the G3 set. }
\resizebox{\columnwidth}{!}{
\begin{tabular}{c   c c c}
\hline
\multicolumn{4}{c}{\textbf{Reference Comparison}}\\
ML models &closed shell&open shell&(G3)\\
&(MAE)&(MAE)&(MAE)\\
\hline
Combined&7.80 &5.17&6.48\\
Combined-LO&13.24 &7.26&10.25\\
Spin--scaled&8.55&3.73&6.14\\
Spin-scaled-LO&11.14 &6.35&8.75\\
\hline
 \end{tabular}
 \label{table:ref}
 }
\end{table}

\begin{figure*}[t!] 
\includegraphics[width=0.98\linewidth]{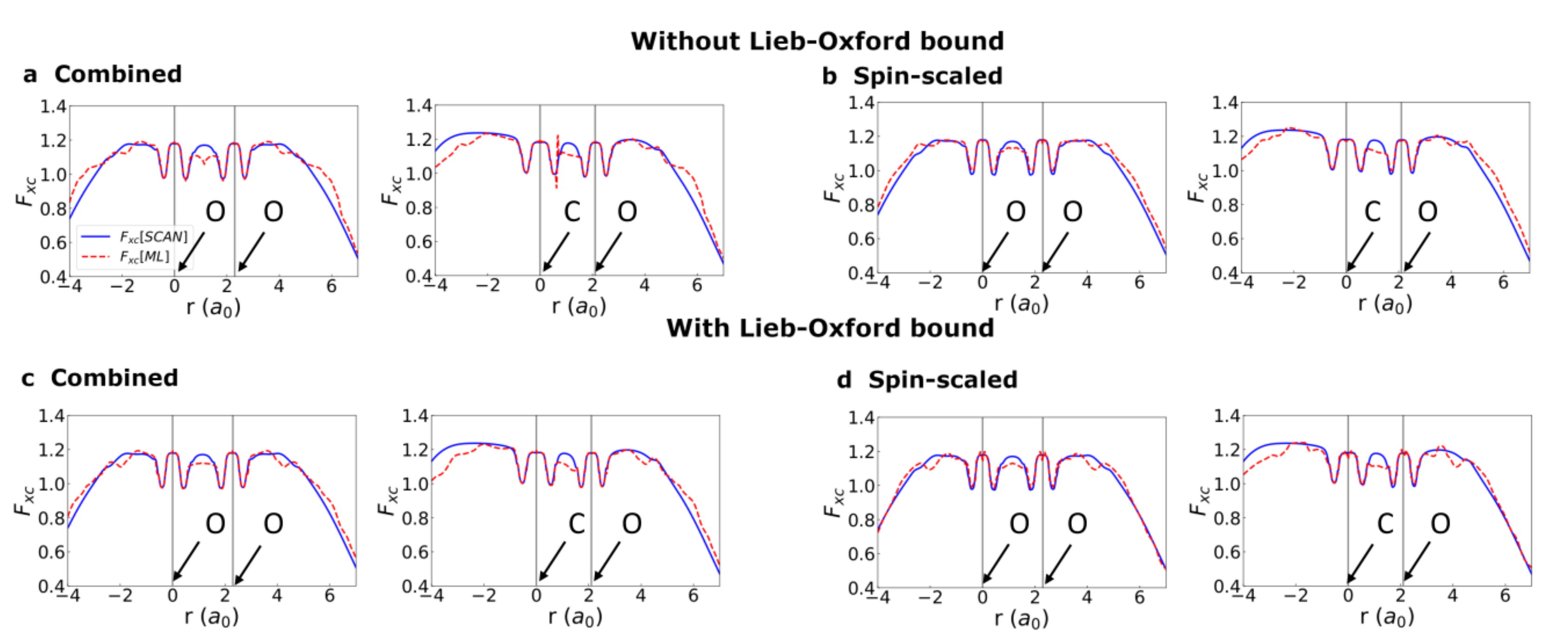} 
\caption{Exchange-correlation enhancement plots for O$_{\mathrm{2}}$ (open shell) and CO (closed shell) from $(\textbf{a})$ combined model and $(\textbf{b})$ spin-scaled model without Lieb--Oxford bound. $(\textbf{c})$, $(\textbf{d})$ are same as $(\textbf{a})$ and $(\textbf{b})$ but for models including Lieb--Oxford bound. The gray vertical lines mark atomic positions.}
\label{fig:mol}
\end{figure*}

The mean absolute error in atomization energy for the G3 set are summarized in Tables \ref{table:scan} and \ref{table:ref} for open-shell, closed-shell, and total collections. Table \ref{table:scan} shows errors relative to total energies calculated using the parent SCAN functional to illustrate model fidelity. Table \ref{table:ref} shows errors relative to standard reference total energies for the set to illustrate the models' chemical accuracy. Overall all models showed comparable performance giving accuracy close to the SCAN functional (MAE 6.53 kcal mol$^{-1}$ ). In particular, enforcing the correct spin-scaling relation in the ML models improved accuracy for the open-shell systems, though this was accompanied by a small reduction in accuracy for the closed shell systems. Introduction of the Lieb--Oxford bound to the combined model deteriorated performance for both open and closed shell systems. This unexpected poor performance suggests that enforcing the Lieb--Oxford bound through Eq. \ref{eq:lb} does not seem to be a successful strategy. The performance of Eq. \ref{eq:lo_x} is more successful, but remains less accurate than the unconstrained equivalent.

It is possible that the reduced performance found is a result of our locally imposing the Lieb--Oxford bound on the model. As the Lieb--Oxford bound is a constraint defined for the total energy imposing it as a constraint on the integrated energy may be more successful, though this would require a significantly different training scheme.

While such post-processing is theoretically convenient, the results from the G3 set and lattice constant below show that it significantly limited model learning during training. We understand this as an effect of the non-linear normalization of Eqs. \ref{eq:lb} and \ref{eq:lo_x} requiring the raw ANN output, $\mathrm{ANN(\mathbf{r})}$, to be simultaneously very large when the target $F_\mathrm{xc}(F_\mathrm{x})$ is small, and very small in energetically significant regions where $F_\mathrm{xc}(F_\mathrm{x})$ approaches $2.215$ (or equivalently $1.174$ for spin-scaled models). The spin-scaled model performs better than the combined model because the number to return (1.174) is relatively smaller compared to 2.215 for combined model and hence error is smaller. An alternative solution for enforcing this constraint could be to simply truncate the range of the network output,
\begin{equation}
    F_\mathrm{xc}^\mathrm{ML} = \max(0, \min(2.215, \mathrm{ANN}(r))),
\end{equation}
and 
\begin{equation}
    F_\mathrm{x}^\mathrm{ML} = \max(0, \min(1.174, \mathrm{ANN}(r))),
\end{equation}
however this may introduce undesirable discontinuities in the partial derivatives of the model, transferring into a non-physical XC potential.

In order to better understand how faithfully the ML models are reproducing the SCAN functional, Figure \ref{fig:mol} compares the XC enhancement factor, $F_\mathrm{xc}$, for the open shell O$_{\mathrm{2}}$ and closed shell CO molecules. We see that all the models are accurate in the areas immediately around the nuclei. This is expected from the good atomic performance as these core regions are relatively unchanged by the covalent bonding. The models deviate more severely around the bond center, with all models underestimating $F_\mathrm{xc}^\mathrm{SCAN}$. In these regions $|\nabla n| \to 0$ and hence $s \to 0$, while the density $n$ remains significant. Such regions are under-represented in the training set, appearing only in small regions at the center of the compressed Ar$_2$ diatomic. This suggests that the spin scaling relation enforced in the spin-scaled model as well as the Lieb--Oxford bound are insufficient to transfer learning from a training set that does not include chemical bonding, onto systems that are chemically bound. Further constraint satisfaction, or inclusion of bonding data into the training set, is likely necessary to improve model accuracy at these important points. We also note that the combined model with no Lieb--Oxford bound exhibits sharp spikes in $F_\mathrm{xc}$ in the bonding regions that are not seen for other ML models or the SCAN functional.

\begin{figure}[t!] 
\includegraphics[width=0.9\linewidth]{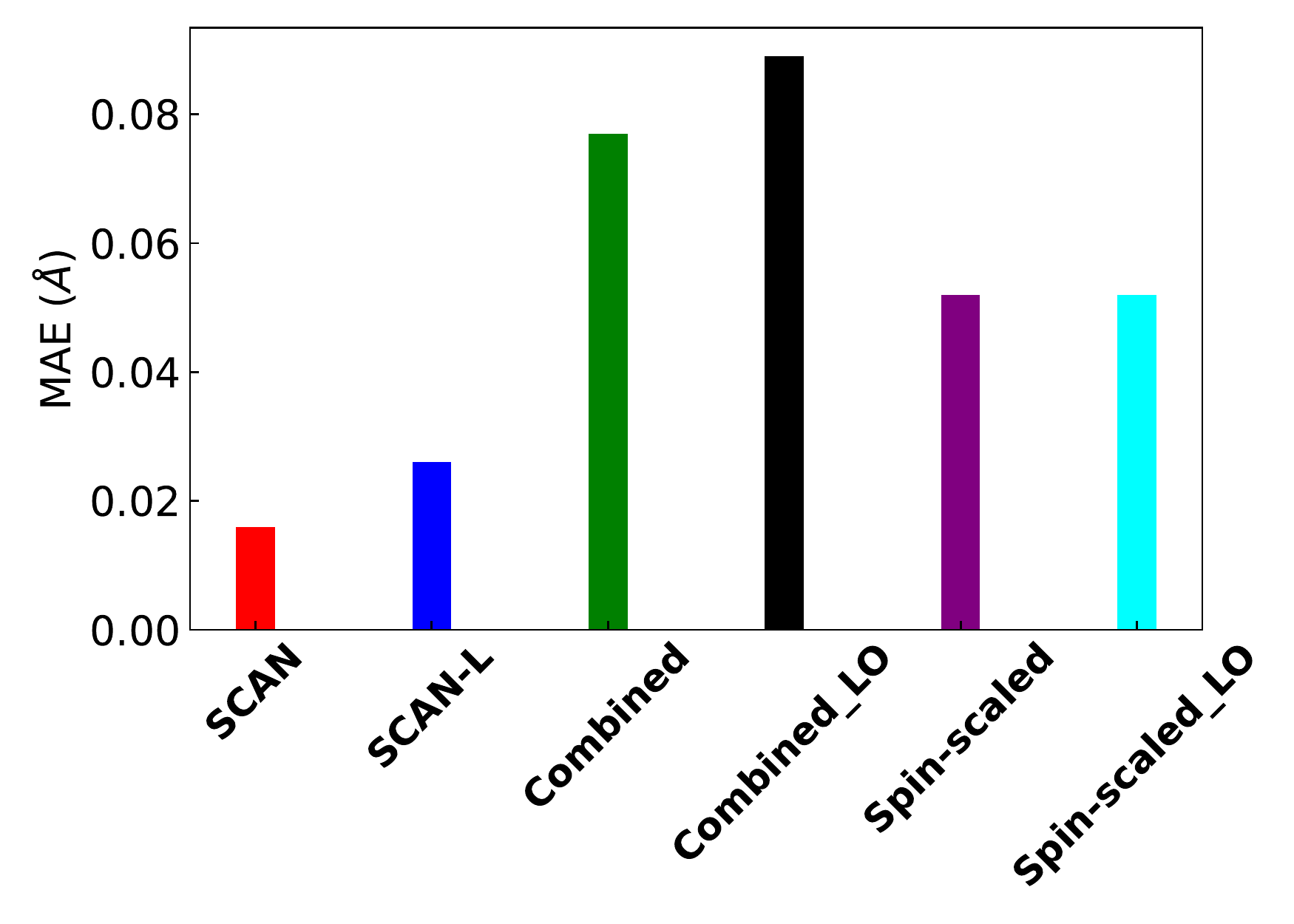} 
\caption{Mean Absolute Error in predicted lattice constant} (MAE, \AA) for SCAN, whose results are obtained from Ref.\cite{sun2015strongly}, SCAN-L, whose results are obtained from Ref. \cite{mejia2018deorbitalized}, combined, combined with Lieb--Oxford bound (denoted ``LO'' within figure), spin-scaled and spin-scaled with Lieb--Oxford bound for the LC20 set of 20 solids with lattice constants ranging from 3.415 to 6.042 \AA \cite{Sun2011}. The ML calculations were performed using fixed self-consistent PBE densities generated using FHI-aims \cite{blum2009ab}.
\label{fig:lat}
\end{figure}

\subsection*{Lattice Constants of solids}

\begin{figure*}[t!] 
\includegraphics[width=0.85\linewidth]{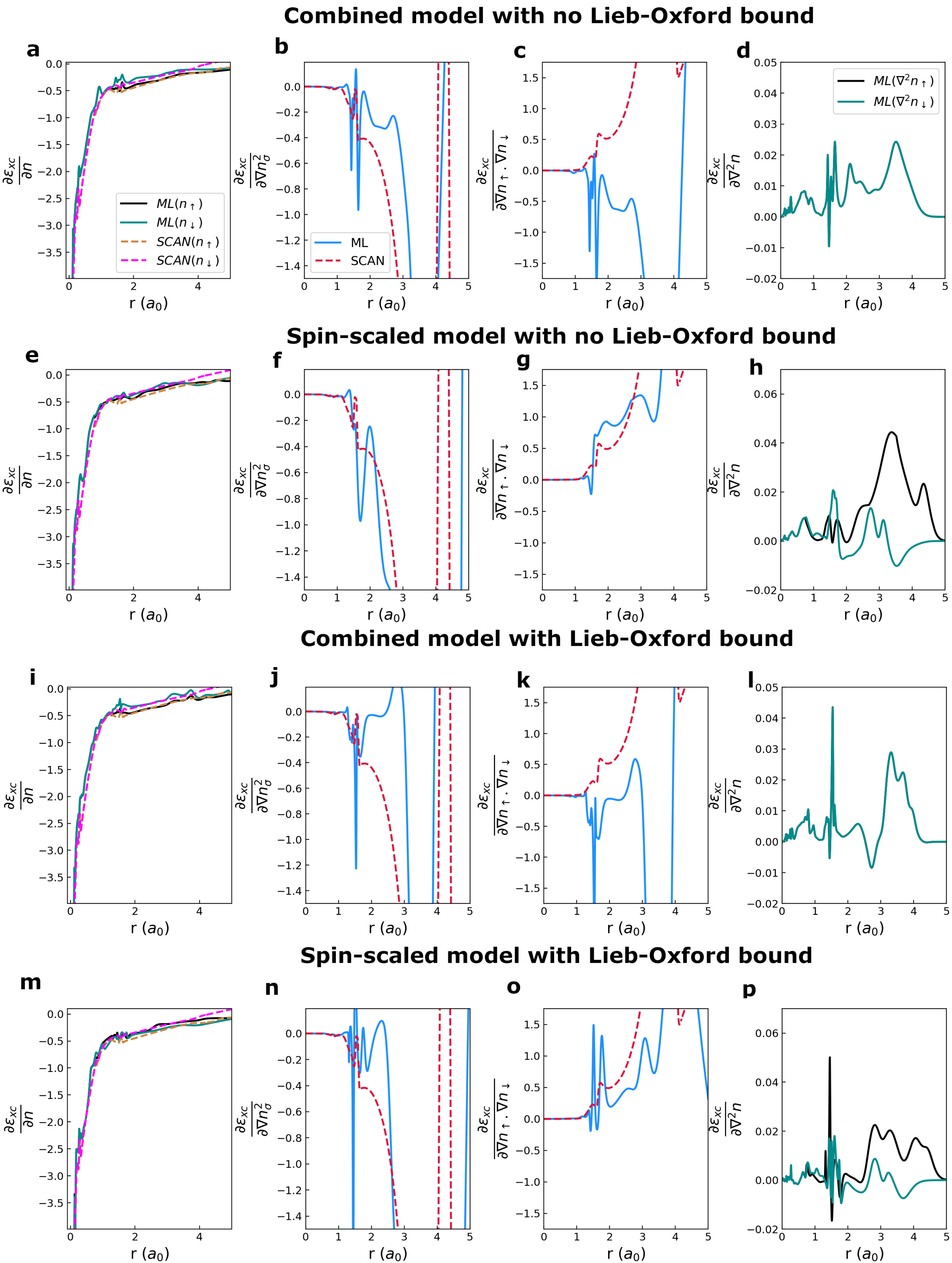} 
\caption{Derivatives of SCAN and ML model XC energy density (combined and spin-scaled) with respect to input ingredients for the silicon atom (not in training set). ML derivatives are represented by solid lines while SCAN are dotted.} 
\label{fig:pot}
\end{figure*}

The transferability of the ML functionals was further tested by calculating the lattice constants of 20 solids from the LC20 test set\cite{Sun2011}. This tests the model's ability to generalise into further unseen chemical environments in periodic systems, as well as requiring description of the energy as a function of nuclear displacement. The LC20 set is therefore a sensitive test of model transferability. The equilibrium lattice constants were determined by a  nine-point fit of total energy per unit cell to the Vinet equation of state around the SCAN equilibrium unit cell volume ($V_{0}$) in a range of $V_{0} \pm 10\%$ \cite{vinet1986universal, vinet1989universal}. 

Figure \ref{fig:lat} compares LC20 results from ML models with the SCAN and SCAN-L functionals. The SCAN results for lattice constants are taken from Ref. \cite{sun2015strongly} and SCAN-L results from Ref. \cite{mejia2018deorbitalized}. The ML calculations were obtained non self-consistently using PBE densities, not SCAN densities as was used for molecular calculations. This is because SCAN suffered numerical instabilies in calculations using FHI-aims\cite{blum2009ab}, with some systems failing to converge. FHI-aims was used for periodic calculations rather than VASP (which was used for Ref. \citenum{sun2015strongly} and \citenum{mejia2018deorbitalized}) as all electron calculations were required to get usable input data for lattice constant calculations. We should therefore be aware of a possible density driven error for the ML models resulting from the PBE densities on which they were evaluated, which could be significant.

The results of Figure \ref{fig:lat} show that the ML models do not perform as well for solids as they do for gas phase atoms and molecules, suggesting difficulty in generalizing knowledge into this untrained domain. The spin-scaled model preformed significantly better than the combined model both with and without enforcing the Lieb--Oxford bound. While the performance of ML models is poor compared to SCAN and SCAN-L, they improve slightly on the PBE GGA \cite{Perdew1996} (MAE 0.060\AA). Having said that, it suggests a similar conclusion to the molecular tests; that the spin-scaling and the Lieb--Oxford bound are insufficient to transfer learning into the periodic systems. Inclusion of such information in the training set is likely necessary to improve model accuracy.

\subsection*{Exchange-correlation potential}
The XC potential is defined as,
\begin{equation}
v_{\mathrm{xc}}(\mathbf{r}) = \frac{\delta E_{\mathrm{xc}}}{\delta n(\mathbf{r})}.
\end{equation}
This is constructed for the ML models from the partial derivatives of the model with respect to its ingredients,
\begin{equation}
\left\{\frac {\partial E_{\mathrm{xc}}^{\mathrm{ML}}}{\partial n(\mathbf{r})},   \frac {\partial E_{\mathrm{xc}}^{\mathrm{ML}}}{\partial |\nabla n(\mathbf{r})|}, \frac {\partial E_{\mathrm{xc}}^{\mathrm{ML}}}{\partial \nabla^{2} n(\mathbf{r})}\right\},
\end{equation}
through repeated application of the chain rule. In practice this is achieved using the back-propagation mechanism of the machine learning framework \cite{liao2019differentiable}. Here the choice of the activation functions for the neuron layers is critically important for obtaining a smooth XC potential appropriate for SCF applications. When the activation functions are differentiated during back-propagation the use of activation functions with discontinuous derivatives, such as the popular rectified linear unit (relu) \cite{brownlee2019gentle}, may introduce discontinuities into the XC potential that can harm SCF convergence and computational efficiency. We therefore used only smooth sigmoid, tanh and elu activation functions within the present models.

Figure \ref{fig:pot} shows partial derivatives of the combined and spin-scaled models with respect to input ingredients, compared against the equivalent for the SCAN functional where they exist. Note that $\epsilon_\mathrm{xc}^{\mathrm{SCAN}}(n, |\nabla n|, \tau)$ and $\epsilon_\mathrm{xc}^{\mathrm{ML}}(n, |\nabla n|, \nabla^2n)$ are necessarily different functions, even though $\epsilon_\mathrm{xc}^{\mathrm{ML}}$ has been trained to reproduce $\epsilon_\mathrm{xc}^{\mathrm{SCAN}}$. Hence, we should not expect their partial derivatives in $n$ and $|\nabla n|$ to match. Figure \ref{fig:pot} ($\textbf{a-d}$) shows the combined model partial derivatives for the test silicon atom, while ($\textbf{e-h}$) shows the same for the spin-scaled model. 

Figures \ref{fig:pot} ($\textbf{a}$), ($\textbf{e}$), ($\textbf{i}$) and ($\textbf{m}$) show that the density partial derivative is comparable to SCAN for both models. Both models exhibit oscillations in this derivative, though these are less severe for the spin-scaled models than the combined models. Figures \ref{fig:pot} ($\textbf{b-c}$) and ($\textbf{j-k}$) for the combined model without and with Lieb--Oxford bound constraint and ($\textbf{f-g}$) and ($\textbf{n-o}$) for spin-scaled model, again with and without Lieb--Oxford bound, show the partial derivatives with respect to the same spin and cross spin gradient components. While the combined model does not distinguish between different spins $\partial \epsilon_{\mathrm{xc}^\mathrm{ML}}/\partial\nabla n_\sigma^2 = \partial \epsilon_{\mathrm{xc}^\mathrm{ML}}/\partial \nabla n_\uparrow\cdot\nabla n_\downarrow$, the spin-scaled model is able to distinguish spin channels so the same-spin and cross-spin partial derivatives are different.  Finally, Figures ($\textbf{d}$), ($\textbf{h}$), ($\textbf{i}$) and ($\textbf{k}$)show the Laplacian partial derivatives. 

It is not clear to what degree the oscillations of the present partial derivatives may affect the SCF performance of the ML models without performing such calculations, which is beyond the scope of the present study. We can reasonably expect the smoother spin-scaled model to outperform the combined model in this regard however, as a result of the reduced oscillations seen in Figure \ref{fig:pot}. The harmful impact of such oscillations may be significantly damped if the regions they occur in are not energetically important. We note that the SCAN functional is known to have problematic oscillations in its XC potential which reduce its computational efficiency but generally do not prevent SCF convergence \cite{Yang2016, Yao2017, Yamamoto2019, Bartok2019, furness2020accurate}. The similarity between the ML and SCAN partial derivatives of Figure \ref{fig:pot} therefore suggest that self-consistency can likely be reached, though this may be sensitive to the choice of starting guess orbitals if they are different from the converged orbitals used in training.

\section*{Discussion}

A broad goal of this work has been to explore how exact constraint adherence can enhance the transferability of ML density functionals beyond a limited set of training data for which their accuracy is theoretically well justified. For the semi-local meta-GGA functionals these are systems for which the XC hole remains well localized (atoms and compressed non-bonding diatoms), or for which the long-range parts of the X and C holes cancel (jellium) \cite{Pitarke2003, Langreth1977}. It is unclear to what extent our models' overall fidelity to SCAN is a limitation of this restricted training data, and to what extent it reflects a fundamental difficulty in representing SCAN from orbital free ingredients. The success of the SCAN-L and related functionals \cite{mejia2017deorbitalization} suggests the former, though repeating the present study with a significantly expanded training set that covers more of the chemical space could provide compelling evidence one way or the other, though is beyond the present scope.

\section*{Conclusion}

We have explored how a philosophy of exact constraints and appropriate norms can be combined with ML techniques in functional design. We have shown a simple test of this idea as a de-orbitalization of the SCAN functional, using the density Laplacian, $\nabla^2{n(\mathbf{r})}$, in place of kinetic energy density $\tau(\mathbf{r})$. Four ML functionals were developed, enforcing a spin-scaling constraint, the Lieb--Oxford bound, both, and neither. These models were trained using a dataset with no chemical bonding, preferring the norms appropriate for semi-local functionals \cite{sun2015strongly}. The model satisfying both the spin-scaling constraint and the Lieb--Oxford bound generally achieved a more balanced performance across the properties tested, though performance was worse than that achieved by the analytical SCAN-L deorbitalization for solids. Given that model performance was generally improved when both constraints were imposed, it is reasonable to believe that engineering in further constraints can enhance robustness of the ML models out of the training domain.

\section*{Data Availability}
The data that supports the finding of this study are available from the corresponding author upon request. Parameters for the ML models presented are provided as supplemental Tensorflow metagraph parameter files.

\section*{Acknowledgments}

J.F. thanks Michael Withnall for his useful discussion and guidance on ML techniques. J.F., K.P., and J.S. acknowledge discussions with John Perdew on semi-local XC functionals. J.S. thanks Kieron Burke for his useful discussion on ML techniques. This work was partially supported by the Research Council of Norway through its Centres of Excellence scheme, project number 262695. We acknowledge financial support from the European Research Council under H2020/ERC Consolidator Grant top DFT (Grant No. 772259).

\section*{Author contributions}
J.S and J.F. designed the project. K.P., J.F., Y.Y., V.B., T.I., and A.T. developed the computational methods, K.P. performed all calculations. K.P., J.F., and J.S. analyzed the data and wrote the manuscript.

\section*{competing interests}
The authors declare no competing interests.

\clearpage{}
\bibliographystyle{ieeetr}{}
\bibliography{reference, SCANML}

\begin{thebibliography}{10}

\bibitem{hohenberg1964inhomogeneous}
P.~Hohenberg and W.~Kohn, ``Inhomogeneous electron gas,'' {\em Physical
  review}, vol.~136, no.~3B, p.~B864, 1964.

\bibitem{kohn1965self}
W.~Kohn and L.~J. Sham, ``Self-consistent equations including exchange and
  correlation effects,'' {\em Physical review}, vol.~140, no.~4A, p.~A1133,
  1965.

\bibitem{perdew2001jacob}
J.~P. Perdew and K.~Schmidt, ``Jacob’s ladder of density functional
  approximations for the exchange-correlation energy,'' in {\em AIP Conference
  Proceedings}, vol.~577, pp.~1--20, American Institute of Physics, 2001.

\bibitem{Sun2013a}
J.~Sun, B.~Xiao, Y.~Fang, R.~Haunschild, P.~Hao, A.~Ruzsinszky, G.~I. Csonka,
  G.~E. Scuseria, and J.~P. Perdew, ``{Density functionals that recognize
  covalent, metallic, and weak bonds},'' {\em Phys. Rev. Lett.}, vol.~111,
  no.~10, p.~106401, 2013.

\bibitem{sun2015strongly}
J.~Sun, A.~Ruzsinszky, and J.~P. Perdew, ``Strongly constrained and
  appropriately normed semilocal density functional,'' {\em Physical review
  letters}, vol.~115, no.~3, p.~036402, 2015.

\bibitem{furness2019enhancing}
J.~W. Furness and J.~Sun, ``Enhancing the efficiency of density functionals
  with an improved iso-orbital indicator,'' {\em Physical Review B}, vol.~99,
  no.~4, p.~041119, 2019.

\bibitem{becke1998new}
A.~D. Becke, ``A new inhomogeneity parameter in density-functional theory,''
  {\em The Journal of chemical physics}, vol.~109, no.~6, pp.~2092--2098, 1998.

\bibitem{gorling1994exact}
A.~G{\"o}rling and M.~Levy, ``Exact kohn-sham scheme based on perturbation
  theory,'' {\em Physical Review A}, vol.~50, no.~1, p.~196, 1994.

\bibitem{wu2003algebraic}
Q.~Wu and W.~Yang, ``Algebraic equation and iterative optimization for the
  optimized effective potential in density functional theory,'' {\em Journal of
  Theoretical and Computational Chemistry}, vol.~2, no.~04, pp.~627--638, 2003.

\bibitem{Seidl1996}
A.~Seidl, A.~G{\"{o}}rling, P.~Vogl, J.~Majewski, and M.~Levy, ``{Generalized
  Kohn-Sham schemes and the band-gap problem},'' {\em Phys. Rev. B}, vol.~53,
  no.~7, pp.~3764--3774, 1996.

\bibitem{Neumann1996}
R.~Neumann, R.~H. Nobes, and N.~C. Handy, ``{Exchange functionals and
  potentials},'' {\em Mol. Phys.}, vol.~87, no.~1, pp.~1--36, 1996.

\bibitem{wu2003direct}
Q.~Wu and W.~Yang, ``A direct optimization method for calculating density
  functionals and exchange--correlation potentials from electron densities,''
  {\em The Journal of chemical physics}, vol.~118, no.~6, pp.~2498--2509, 2003.

\bibitem{mejia2017deorbitalization}
D.~Mejia-Rodriguez and S.~Trickey, ``Deorbitalization strategies for
  meta-generalized-gradient-approximation exchange-correlation functionals,''
  {\em Physical Review A}, vol.~96, no.~5, p.~052512, 2017.

\bibitem{mejia2018deorbitalized}
D.~Mejia-Rodriguez and S.~Trickey, ``Deorbitalized meta-gga
  exchange-correlation functionals in solids,'' {\em Physical Review B},
  vol.~98, no.~11, p.~115161, 2018.

\bibitem{della2016kinetic}
F.~Della~Sala, E.~Fabiano, and L.~A. Constantin, ``Kinetic-energy-density
  dependent semilocal exchange-correlation functionals,'' {\em International
  Journal of Quantum Chemistry}, vol.~116, no.~22, pp.~1641--1694, 2016.

\bibitem{dorigo1993genetics}
M.~Dorigo and U.~Schnepf, ``Genetics-based machine learning and behavior-based
  robotics: a new synthesis,'' {\em IEEE Transactions on Systems, Man, and
  Cybernetics}, vol.~23, no.~1, pp.~141--154, 1993.

\bibitem{salichs2006maggie}
M.~A. Salichs, R.~Barber, A.~M. Khamis, M.~Malfaz, J.~F. Gorostiza, R.~Pacheco,
  R.~Rivas, A.~Corrales, E.~Delgado, and D.~Garcia, ``Maggie: A robotic
  platform for human-robot social interaction,'' in {\em 2006 IEEE conference
  on robotics, automation and mechatronics}, pp.~1--7, IEEE, 2006.

\bibitem{macleod2010time}
N.~MacLeod, M.~Benfield, and P.~Culverhouse, ``Time to automate
  identification,'' {\em Nature}, vol.~467, no.~7312, pp.~154--155, 2010.

\bibitem{chittka2012your}
L.~Chittka and A.~Dyer, ``Your face looks familiar,'' {\em Nature}, vol.~481,
  no.~7380, pp.~154--155, 2012.

\bibitem{ashley2015precision}
E.~A. Ashley, ``The precision medicine initiative: a new national effort,''
  {\em Jama}, vol.~313, no.~21, pp.~2119--2120, 2015.

\bibitem{schatz2013dna}
M.~C. Schatz and B.~Langmead, ``The dna data deluge,'' {\em Ieee Spectrum},
  vol.~50, no.~7, pp.~28--33, 2013.

\bibitem{ding2014similarity}
H.~Ding, I.~Takigawa, H.~Mamitsuka, and S.~Zhu, ``Similarity-based machine
  learning methods for predicting drug--target interactions: a brief review,''
  {\em Briefings in bioinformatics}, vol.~15, no.~5, pp.~734--747, 2014.

\bibitem{silver2016mastering}
D.~Silver, A.~Huang, C.~J. Maddison, A.~Guez, L.~Sifre, G.~Van Den~Driessche,
  J.~Schrittwieser, I.~Antonoglou, V.~Panneershelvam, M.~Lanctot, {\em et~al.},
  ``Mastering the game of go with deep neural networks and tree search,'' {\em
  nature}, vol.~529, no.~7587, pp.~484--489, 2016.

\bibitem{snyder2012finding}
J.~C. Snyder, M.~Rupp, K.~Hansen, K.-R. M{\"u}ller, and K.~Burke, ``Finding
  density functionals with machine learning,'' {\em Physical review letters},
  vol.~108, no.~25, p.~253002, 2012.

\bibitem{snyder2013orbital}
J.~C. Snyder, M.~Rupp, K.~Hansen, L.~Blooston, K.-R. M{\"u}ller, and K.~Burke,
  ``Orbital-free bond breaking via machine learning,'' {\em The Journal of
  chemical physics}, vol.~139, no.~22, p.~224104, 2013.

\bibitem{brockherde2017bypassing}
F.~Brockherde, L.~Vogt, L.~Li, M.~E. Tuckerman, K.~Burke, and K.-R. M{\"u}ller,
  ``Bypassing the kohn-sham equations with machine learning,'' {\em Nature
  communications}, vol.~8, no.~1, pp.~1--10, 2017.

\bibitem{liu2017improving}
Q.~Liu, J.~Wang, P.~Du, L.~Hu, X.~Zheng, and G.~Chen, ``Improving the
  performance of long-range-corrected exchange-correlation functional with an
  embedded neural network,'' {\em The Journal of Physical Chemistry A},
  vol.~121, no.~38, pp.~7273--7281, 2017.

\bibitem{nagai2018neural}
R.~Nagai, R.~Akashi, S.~Sasaki, and S.~Tsuneyuki, ``Neural-network kohn-sham
  exchange-correlation potential and its out-of-training transferability,''
  {\em The Journal of chemical physics}, vol.~148, no.~24, p.~241737, 2018.

\bibitem{wellendorff2012density}
J.~Wellendorff, K.~T. Lundgaard, A.~M{\o}gelh{\o}j, V.~Petzold, D.~D. Landis,
  J.~K. N{\o}rskov, T.~Bligaard, and K.~W. Jacobsen, ``Density functionals for
  surface science: Exchange-correlation model development with bayesian error
  estimation,'' {\em Physical Review B}, vol.~85, no.~23, p.~235149, 2012.

\bibitem{li2016pure}
L.~Li, T.~E. Baker, S.~R. White, K.~Burke, {\em et~al.}, ``Pure density
  functional for strong correlation and the thermodynamic limit from machine
  learning,'' {\em Physical Review B}, vol.~94, no.~24, p.~245129, 2016.

\bibitem{westermayr2021perspective}
J.~Westermayr, M.~Gastegger, K.~T. Sch{\"u}tt, and R.~J. Maurer, ``Perspective
  on integrating machine learning into computational chemistry and materials
  science,'' {\em arXiv preprint arXiv:2102.08435}, 2021.

\bibitem{sun2016accurate}
J.~Sun, R.~C. Remsing, Y.~Zhang, Z.~Sun, A.~Ruzsinszky, H.~Peng, Z.~Yang,
  A.~Paul, U.~Waghmare, X.~Wu, {\em et~al.}, ``Accurate first-principles
  structures and energies of diversely bonded systems from an efficient density
  functional,'' {\em Nature chemistry}, vol.~8, no.~9, p.~831, 2016.

\bibitem{chen2017ab}
M.~Chen, H.-Y. Ko, R.~C. Remsing, M.~F.~C. Andrade, B.~Santra, Z.~Sun,
  A.~Selloni, R.~Car, M.~L. Klein, J.~P. Perdew, {\em et~al.}, ``Ab initio
  theory and modeling of water,'' {\em Proceedings of the National Academy of
  Sciences}, vol.~114, no.~41, pp.~10846--10851, 2017.

\bibitem{remsing2017dependence}
R.~C. Remsing, M.~L. Klein, and J.~Sun, ``Dependence of the structure and
  dynamics of liquid silicon on the choice of density functional
  approximation,'' {\em Physical Review B}, vol.~96, no.~2, p.~024203, 2017.

\bibitem{gautam2018evaluating}
G.~S. Gautam and E.~A. Carter, ``Evaluating transition metal oxides within
  dft-scan and scan+ u frameworks for solar thermochemical applications,'' {\em
  Physical Review Materials}, vol.~2, no.~9, p.~095401, 2018.

\bibitem{furness2018accurate}
J.~W. Furness, Y.~Zhang, C.~Lane, I.~G. Buda, B.~Barbiellini, R.~S. Markiewicz,
  A.~Bansil, and J.~Sun, ``An accurate first-principles treatment of
  doping-dependent electronic structure of high-temperature cuprate
  superconductors,'' {\em Communications Physics}, vol.~1, no.~1, pp.~1--6,
  2018.

\bibitem{lane2018antiferromagnetic}
C.~Lane, J.~W. Furness, I.~G. Buda, Y.~Zhang, R.~S. Markiewicz, B.~Barbiellini,
  J.~Sun, and A.~Bansil, ``Antiferromagnetic ground state of la 2 cuo 4: A
  parameter-free ab initio description,'' {\em Physical Review B}, vol.~98,
  no.~12, p.~125140, 2018.

\bibitem{pokharel2020sensitivity}
K.~Pokharel, C.~Lane, J.~W. Furness, R.~Zhang, J.~Ning, B.~Barbiellini, R.~S.
  Markiewicz, Y.~Zhang, A.~Bansil, and J.~Sun, ``Sensitivity of the electronic
  and magnetic structures of high-temperature cuprate superconductors to
  exchange-correlation density functionals,'' {\em arXiv preprint
  arXiv:2004.08047}, 2020.

\bibitem{hollingsworth2018can}
J.~Hollingsworth, L.~Li, T.~E. Baker, and K.~Burke, ``Can exact conditions
  improve machine-learned density functionals?,'' {\em The Journal of chemical
  physics}, vol.~148, no.~24, p.~241743, 2018.

\bibitem{nagai2022machine}
R.~Nagai, R.~Akashi, and O.~Sugino, ``Machine-learning-based exchange
  correlation functional with physical asymptotic constraints,'' {\em Physical
  Review Research}, vol.~4, no.~1, p.~013106, 2022.

\bibitem{kirkpatrick2021pushing}
J.~Kirkpatrick, B.~McMorrow, D.~H. Turban, A.~L. Gaunt, J.~S. Spencer, A.~G.
  Matthews, A.~Obika, L.~Thiry, M.~Fortunato, D.~Pfau, {\em et~al.}, ``Pushing
  the frontiers of density functionals by solving the fractional electron
  problem,'' {\em Science}, vol.~374, no.~6573, pp.~1385--1389, 2021.

\bibitem{lieb1981improved}
E.~H. Lieb and S.~Oxford, ``Improved lower bound on the indirect coulomb
  energy,'' {\em International Journal of Quantum Chemistry}, vol.~19, no.~3,
  pp.~427--439, 1981.

\bibitem{oliver1979spin}
G.~Oliver and J.~Perdew, ``Spin-density gradient expansion for the kinetic
  energy,'' {\em Physical Review A}, vol.~20, no.~2, p.~397, 1979.

\bibitem{levy1985hellmann}
M.~Levy and J.~P. Perdew, ``Hellmann-feynman, virial, and scaling requisites
  for the exact universal density functionals. shape of the correlation
  potential and diamagnetic susceptibility for atoms,'' {\em Physical Review
  A}, vol.~32, no.~4, p.~2010, 1985.

\bibitem{svendsen1996gradient}
P.-S. Svendsen and U.~von Barth, ``Gradient expansion of the exchange energy
  from second-order density response theory,'' {\em Physical Review B},
  vol.~54, no.~24, p.~17402, 1996.

\bibitem{pollack2000evaluating}
L.~Pollack and J.~Perdew, ``Evaluating density functional performance for the
  quasi-two-dimensional electron gas,'' {\em Journal of Physics: Condensed
  Matter}, vol.~12, no.~7, p.~1239, 2000.

\bibitem{perdew1992accurate}
J.~P. Perdew and Y.~Wang, ``Accurate and simple analytic representation of the
  electron-gas correlation energy,'' {\em Physical review B}, vol.~45, no.~23,
  p.~13244, 1992.

\bibitem{perdew2014gedanken}
J.~P. Perdew, A.~Ruzsinszky, J.~Sun, and K.~Burke, ``Gedanken densities and
  exact constraints in density functional theory,'' {\em The Journal of
  chemical physics}, vol.~140, no.~18, p.~18A533, 2014.

\bibitem{perdew2018erratum}
J.~P. Perdew and Y.~Wang, ``Erratum: Accurate and simple analytic
  representation of the electron-gas correlation energy [phys. rev. b 45, 13244
  (1992)],'' {\em Physical Review B}, vol.~98, no.~7, p.~079904, 2018.

\bibitem{Perdew1996}
J.~P. Perdew, K.~Burke, and M.~Ernzerhof, ``{Generalized Gradient Approximation
  Made Simple.},'' {\em Phys. Rev. Lett.}, vol.~77, no.~18, pp.~3865--3868,
  1996.

\bibitem{perdew1991electronic}
J.~P. Perdew, P.~Ziesche, and H.~Eschrig, ``Electronic structure of solids’
  91,'' 1991.

\bibitem{seidl2000simulation}
M.~Seidl, J.~P. Perdew, and S.~Kurth, ``Simulation of all-order
  density-functional perturbation theory, using the second order and the
  strong-correlation limit,'' {\em Physical review letters}, vol.~84, no.~22,
  p.~5070, 2000.

\bibitem{tao2003climbing}
J.~Tao, J.~P. Perdew, V.~N. Staroverov, and G.~E. Scuseria, ``Climbing the
  density functional ladder: Nonempirical meta--generalized gradient
  approximation designed for molecules and solids,'' {\em Physical Review
  Letters}, vol.~91, no.~14, p.~146401, 2003.

\bibitem{tao2008nonempirical}
J.~Tao, J.~P. Perdew, L.~M. Almeida, C.~Fiolhais, and S.~K{\"u}mmel,
  ``Nonempirical density functionals investigated for jellium: Spin-polarized
  surfaces, spherical clusters, and bulk linear response,'' {\em Physical
  Review B}, vol.~77, no.~24, p.~245107, 2008.

\bibitem{kalman1992tanh}
B.~L. Kalman and S.~C. Kwasny, ``Why tanh: choosing a sigmoidal function,'' in
  {\em [Proceedings 1992] IJCNN International Joint Conference on Neural
  Networks}, vol.~4, pp.~578--581, IEEE, 1992.

\bibitem{Oliver1979}
G.~L. Oliver and J.~P. Perdew, ``{Spin-density gradient expansion for the
  kinetic energy},'' {\em Phys. Rev. A}, vol.~20, no.~2, pp.~397--403, 1979.

\bibitem{Clementi1974}
E.~Clementi and C.~Roetti, ``{Roothaan-Hartree-Fock Atomic Wavefunctions},''
  {\em At. Data Nucl. Data Tables}, vol.~14, no.~3, pp.~177--478, 1974.

\bibitem{Koga1999}
T.~Koga, K.~Kanayama, S.~Watanabe, and A.~J. Thakkar, ``{Analytical
  Hartree-Fock wave functions subject to cusp and asymptotic constraints: He to
  Xe, Li+ to Cs+, H- to I-},'' {\em Int. J. Quantum Chem.}, vol.~71, no.~6,
  pp.~491--497, 1999.

\bibitem{Furness2021a}
J.~W. Furness and S.~Lehtola, ``{Hartree-Fock Orbitals for Spherical Atoms - a
  python toolbox},'' 2021.

\bibitem{Almeida2002}
L.~M. Almeida, J.~P. Perdew, and C.~Fiolhais, ``{Surface and curvature energies
  from jellium spheres: Density functional hierarchy and quantum Monte
  Carlo},'' {\em Phys. Rev. B}, vol.~66, no.~7, p.~075115, 2002.

\bibitem{Wood2007}
B.~Wood, N.~D.~M. Hine, W.~M.~C. Foulkes, and P.~Garc{\'{i}}a-Gonz{\'{a}}lez,
  ``{Quantum Monte Carlo calculations of the surface energy of an electron
  gas},'' {\em Phys. Rev. B}, vol.~76, no.~3, p.~035403, 2007.

\bibitem{abadi2016tensorflow}
M.~Abadi, P.~Barham, J.~Chen, Z.~Chen, A.~Davis, J.~Dean, M.~Devin,
  S.~Ghemawat, G.~Irving, M.~Isard, {\em et~al.}, ``Tensorflow: A system for
  large-scale machine learning,'' in {\em 12th $\{$USENIX$\}$ symposium on
  operating systems design and implementation ($\{$OSDI$\}$ 16)}, pp.~265--283,
  2016.

\bibitem{han1995influence}
J.~Han and C.~Moraga, ``The influence of the sigmoid function parameters on the
  speed of backpropagation learning,'' in {\em International Workshop on
  Artificial Neural Networks}, pp.~195--201, Springer, 1995.

\bibitem{clevert2015fast}
D.-A. Clevert, T.~Unterthiner, and S.~Hochreiter, ``Fast and accurate deep
  network learning by exponential linear units (elus),'' {\em arXiv preprint
  arXiv:1511.07289}, 2015.

\bibitem{scikit-learn}
F.~Pedregosa, G.~Varoquaux, A.~Gramfort, V.~Michel, B.~Thirion, O.~Grisel,
  M.~Blondel, P.~Prettenhofer, R.~Weiss, V.~Dubourg, J.~Vanderplas, A.~Passos,
  D.~Cournapeau, M.~Brucher, M.~Perrot, and E.~Duchesnay, ``Scikit-learn:
  Machine learning in {P}ython,'' {\em Journal of Machine Learning Research},
  vol.~12, pp.~2825--2830, 2011.

\bibitem{kingma2014adam}
D.~P. Kingma and J.~Ba, ``Adam: A method for stochastic optimization,'' {\em
  arXiv preprint arXiv:1412.6980}, 2014.

\bibitem{nair2010rectified}
V.~Nair and G.~E. Hinton, ``Rectified linear units improve restricted boltzmann
  machines,'' in {\em Icml}, 2010.

\bibitem{Curtiss2000}
L.~A. Curtiss, K.~Raghavachari, P.~C. Redfern, and J.~A. Pople, ``{Assessment
  of Gaussian-3 and density functional theories for a larger experimental test
  set},'' {\em J. Chem. Phys.}, vol.~112, no.~17, pp.~7374--7383, 2000.

\bibitem{clark1983efficient}
T.~Clark, J.~Chandrasekhar, G.~W. Spitznagel, and P.~V.~R. Schleyer,
  ``Efficient diffuse function-augmented basis sets for anion calculations.
  iii. the 3-21+ g basis set for first-row elements, li--f,'' {\em Journal of
  Computational Chemistry}, vol.~4, no.~3, pp.~294--301, 1983.

\bibitem{frisch1984self}
M.~J. Frisch, J.~A. Pople, and J.~S. Binkley, ``Self-consistent molecular
  orbital methods 25. supplementary functions for gaussian basis sets,'' {\em
  The Journal of chemical physics}, vol.~80, no.~7, pp.~3265--3269, 1984.

\bibitem{QUEST}
A.~M. Teale, J.~W. Furness, T.~J.~P. Irons, M.~S. Ryley, J.~Zemen, M.~Withnal,
  and J.~Elsdon, ``{QUantum Electronic Structure Techniques (QUEST)},'' 2017.

\bibitem{Sun2011}
J.~Sun, M.~Marsman, G.~I. Csonka, A.~Ruzsinszky, P.~Hao, Y.~S. Kim, G.~Kresse,
  and J.~P. Perdew, ``{Self-consistent meta-generalized gradient approximation
  within the projector-augmented-wave method},'' {\em Phys. Rev. B}, vol.~84,
  no.~3, p.~035117, 2011.

\bibitem{blum2009ab}
V.~Blum, R.~Gehrke, F.~Hanke, P.~Havu, V.~Havu, X.~Ren, K.~Reuter, and
  M.~Scheffler, ``Ab initio molecular simulations with numeric atom-centered
  orbitals,'' {\em Computer Physics Communications}, vol.~180, no.~11,
  pp.~2175--2196, 2009.

\bibitem{vinet1986universal}
P.~Vinet, J.~Ferrante, J.~Smith, and J.~Rose, ``A universal equation of state
  for solids,'' {\em Journal of Physics C: Solid State Physics}, vol.~19,
  no.~20, p.~L467, 1986.

\bibitem{vinet1989universal}
P.~Vinet, J.~H. Rose, J.~Ferrante, and J.~R. Smith, ``Universal features of the
  equation of state of solids,'' {\em Journal of Physics: Condensed Matter},
  vol.~1, no.~11, p.~1941, 1989.

\bibitem{liao2019differentiable}
H.-J. Liao, J.-G. Liu, L.~Wang, and T.~Xiang, ``Differentiable programming
  tensor networks,'' {\em Physical Review X}, vol.~9, no.~3, p.~031041, 2019.

\bibitem{brownlee2019gentle}
J.~Brownlee, ``A gentle introduction to the rectified linear unit (relu),''
  {\em Machine learning mastery}, vol.~6, 2019.

\bibitem{Yang2016}
Z.-h. Yang, H.~Peng, J.~Sun, and J.~P. Perdew, ``{More realistic band gaps from
  meta-generalized gradient approximations: Only in a generalized Kohn-Sham
  scheme},'' {\em Phys. Rev. B}, vol.~93, no.~20, p.~205205, 2016.

\bibitem{Yao2017}
Y.~Yao and Y.~Kanai, ``{Plane-wave pseudopotential implementation and
  performance of SCAN meta-GGA exchange-correlation functional for extended
  systems},'' {\em J. Chem. Phys.}, vol.~146, no.~22, p.~224105, 2017.

\bibitem{Yamamoto2019}
Y.~Yamamoto, C.~M. Diaz, L.~Basurto, K.~A. Jackson, T.~Baruah, and R.~R. Zope,
  ``{Fermi-L{\"{o}}wdin orbital self-interaction correction using the strongly
  constrained and appropriately normed meta-GGA functional},'' {\em J. Chem.
  Phys.}, vol.~151, no.~15, p.~154105, 2019.

\bibitem{Bartok2019}
A.~P. Bart{\'{o}}k and J.~R. Yates, ``{Regularized SCAN functional},'' {\em J.
  Chem. Phys.}, vol.~150, no.~16, p.~161101, 2019.

\bibitem{furness2020accurate}
J.~W. Furness, A.~D. Kaplan, J.~Ning, J.~P. Perdew, and J.~Sun, ``Accurate and
  numerically efficient r2scan meta-generalized gradient approximation,'' {\em
  The journal of physical chemistry letters}, vol.~11, no.~19, pp.~8208--8215,
  2020.

\bibitem{Pitarke2003}
J.~M. Pitarke and J.~P. Perdew, ``Metal surface energy: Persistent cancellation
  of short-range correlation effects beyond the random phase approximation,''
  {\em Phys. Rev. B}, vol.~67, p.~045101, Jan 2003.

\bibitem{Langreth1977}
D.~C. Langreth and J.~P. Perdew, ``Exchange-correlation energy of a metallic
  surface: Wave-vector analysis,'' {\em Phys. Rev. B}, vol.~15, pp.~2884--2901,
  Mar 1977.

\end{thebibliography}
\end{document}


\title{Supplementary materials for ``Exact constraints and appropriate norms in machine learned exchange-correlation functionals''} 
\author
{Kanun Pokharel,$^{1,*}$ James W. Furness,$^{1}$ Yi Yao,$^{2,3}$ Volker Blum$^{*2}$, Tom J.P. Irons,$^{4}$ Andrew M. Teale,$^{*4,5}$ Jianwei Sun$^{*1}$

\textit{\normalsize{$^{1}$Department of Physics and Engineering Physics, Tulane University, New Orleans, LA 70118, USA}}\\
\textit{\normalsize{$^{2}$Thomas Lord Department of Mechanical Engineering and Material Science, Duke University, Durham, North Carolina 27708, USA}}\\
\textit{\normalsize{$^{3}$University of North Carolina at Chapel Hill, Chapel Hill, NC 27599, USA}}\\
\textit{\normalsize{$^{4}$ School of Chemistry, University of Nottingham, Nottingham NG7 2RD, UK}}\\
\textit{\normalsize{$^{5}$ Hylleraas Centre for Quantum Molecular Sciences, Department of Chemistry, University of Oslo, P.O. Box 1033, Blindern, N-0315 Oslo, Norway}}\\ 
\normalsize{$^\ast$To whom correspondence should be addressed; E-mail:  jsun@tulane.edu}
}
\date{\today}
\maketitle
\clearpage{}
\tableofcontents
\clearpage{}

\section{Atomization energy (AE6) results}
\begin{table}[h!]
\caption{Atomization Energies results from the combined and spin-scaled models with and without Lieb-Oxford bounds (LO). The values are compared against the reference values. For completeness, parent SCAN values are also presented. All values are in kcal mol$^{-1}$. All SCAN calculations were performed fully self-consistently with the 6-311++G(3df,3pd) basis set in QUEST. The ML calculations are performed non self-consistently.}
\begin{center}
\begin{tabular}{lc c c c c c c}
\hline\hline
&Molecules&Benchmark & SCAN& Combined & Spin-scaled& Combined-LO& Spin-scaled-LO\\
\hline
&SiH$_{4}$ &322.4& 322.55&317.77&324.90&311.84&334.74\\
&SiO &192.08&186.53 &187.44&196.42&192.08&197.55\\
&S$_{2}$ &101.67&109.43&107.69&107.52&101.67&102.71\\
&C$_{3}$H$_{4}$&704.79&704.04&709.91&707.69&711.90&705.25\\
&C$_{2}$H$_{2}$O$_{2}$ &633.35&633.50 &631.68&627.94&628.37&639.80\\
&C$_{4}$H$_{8}$ &1149.01&1153.65&1143.43&1142.93&1147.01&1150.32\\
\hline\hline
&MAE &-&3.16&4.60&4.51&6.90&4.52\\
 \end{tabular}
\end{center}
 \label{table:ae6}
\end{table}

\section{Barrier height (BH6) results}
\begin{table}[h!]
\caption{Barrier height results from the combined and spin-scaled models with and without Lieb-Oxford bound (LO). The values are compared against the reference values. For completeness, parent SCAN values are also presented. All values are in kcal mol$^{-1}$. All SCAN calculations were performed fully self-consistently with the 6-311++G(3df,3pd) basis set in QUEST. The ML calculations are performed non self-consistently.}
\begin{center}
\resizebox{\columnwidth}{!}{
\begin{tabular}{lc c c c c c c}
\hline\hline
&&Benchmark & SCAN& Combined & Spin-scaled &Combined-LO&Spin-scaled-LO\\
\hline
&Forward TS-H-OH-H$_{2}$ &10.7& 7.47&10.82&9.37&7.47&3.73\\
&Backward TS-H-OH-H$_{2}$&13.1&11.56 &5.56&12.25&5.75&5.74\\
&Forward TS-OH-CH$_{4}$-CH$_{3}$-H$_{2}$O &6.7&8.61&8.57&6.55&5.62&2.11\\
&Backward TS-OH-CH$_{4}$-CH$_{3}$-H$_{2}$O&19.6&7.97&9.79&11.17&7.97&4.28\\
&Forward TS-H-H$_{2}$S-H$_{2}$-SH &3.5&6.23 &10.42&4.70&6.23&0.79\\
&Backward TS-H-H$_{2}$S-H$_{2}$-SH&17.3&6.58&15.07&10.89&6.58&3.88\\
\hline\hline
&MAE &-&8.07&10.04&9.15&7.99&3.42\\
 \end{tabular}
 }
\end{center}
 \label{table:bh6}
\end{table}
\pagebreak

\section{ML models' exchange energy comparison with SCAN exchange energy}
\begin{table}[h!]
\caption{Error in machine learned exchange component of the spin-scaled model with and without Lieb-Oxford bound (LO), shown as total exchange energy energy difference for spherical atom densities relative to SCAN exchange. All values are in Hartree calculated post-SCF from accurate spherical Hartree--Fock Slater-type orbitals \cite{Clementi1974, Koga1999, Furness2021a}.}
\begin{center}
\begin{tabular}{c c c}
\hline
&Unbounded&Bounded\\
\hline
H&-0.001&-0.001\\
He&-0.002&-0.008\\
Li&0.001&-0.017\\
Be&0.006&-0.021\\
B&0.020&-0.009\\
C&0.014&-0.011\\
N&-0.016&-0.035\\
O&0.013&0.011\\
F&0.007&0.012\\
Ne&-0.031&-0.034\\
Na&-0.038&-0.081\\
Mg&-0.047&-0.146\\
Si&-0.062&-0.271\\
P&-0.097&-0.338\\
Ar&-0.165&-0.496\\
K&-0.177&-0.548\\
Cr&-0.309&-0.960\\
Cu&-0.440&-1.681\\
Cu$^+$&-0.428&-1.678\\
As&-0.621&-2.333\\
Kr&-0.739&-2.845\\
Cl&-0.122&-0.418\\
\hline
ME&-0.147&-0.541\\
MAE&0.153&0.543\\
\hline
\end{tabular}
\end{center}
 \label{table:open1}
\end{table}

\begin{longtable}{lrr}
\caption[Error in exchange model]{Error in machine learned exchange component of the spin-scaled model with and without Lieb-Oxford bound (LO), shown as difference in total exchange energy energy for the G3 set relative to SCAN exchange. All values are in Hartree calculated post-SCF from converged SCAN orbitals.}\\
\toprule
                                                         Molecule &  Unbounded &  Bounded \\
\midrule
\endfirsthead
\caption[]{Error in machine learned exchange component of the spin-scaled model with and without Lieb-Oxford bound (LO), shown as difference in total exchange energy energy for the G3 set relative to SCAN exchange. All values are in Hartree calculated post-SCF from converged SCAN orbitals.} \\
\toprule
                                                         Molecule &  Unbounded &  Bounded \\
\midrule
\endhead
\midrule
\multicolumn{3}{r}{{Continued on next page}} \\
\midrule
\endfoot

\bottomrule
\endlastfoot
                                                              LiH &      0.000 &   -0.018 \\
                                                              BeH &     -0.001 &   -0.027 \\
                                                               CH &     -0.002 &   -0.023 \\
                                   CH\mol{2} (\excite{3}B\mol{1}) &     -0.035 &   -0.060 \\
                                   CH\mol{2} (\excite{1}A\mol{1}) &     -0.021 &   -0.043 \\
                                                        CH\mol{3} &     -0.049 &   -0.074 \\
                                              Methane (CH\mol{4}) &     -0.067 &   -0.095 \\
                                                               NH &     -0.017 &   -0.033 \\
                                                        NH\mol{2} &     -0.034 &   -0.046 \\
                                              Ammonia (NH\mol{3}) &     -0.057 &   -0.071 \\
                                                               OH &     -0.018 &   -0.019 \\
                                                Water (H\mol{2}O) &     -0.040 &   -0.041 \\
                                           Hydrogen fluoride (HF) &     -0.029 &   -0.026 \\
                                  SiH\mol{2} (\excite{1}A\mol{1}) &     -0.083 &   -0.299 \\
                                  SiH\mol{2} (\excite{3}B\mol{1}) &     -0.092 &   -0.304 \\
                                                       SiH\mol{3} &     -0.101 &   -0.315 \\
                                              Silane (SiH\mol{4}) &     -0.112 &   -0.329 \\
                                                        PH\mol{2} &     -0.110 &   -0.361 \\
                                                        PH\mol{3} &     -0.126 &   -0.380 \\
                                     Hydrogen sulfide (H\mol{2}S) &     -0.137 &   -0.417 \\
                                          Hydrogen chloride (HCl) &     -0.150 &   -0.457 \\
                                                        Li\mol{2} &      0.005 &   -0.029 \\
                                                              LiF &     -0.023 &   -0.030 \\
                                     Acetylene (C\mol{2}H\mol{2}) &     -0.056 &   -0.114 \\
                                   Ethylene (H\mol{2}C=CH\mol{2}) &     -0.090 &   -0.148 \\
                                     Ethane (H\mol{3}C-CH\mol{3}) &     -0.125 &   -0.187 \\
                                                               CN &     -0.019 &   -0.065 \\
                                           Hydrogen cyanide (HCN) &     -0.042 &   -0.090 \\
                                                               CO &     -0.027 &   -0.063 \\
                                                              HCO &     -0.043 &   -0.080 \\
                                       Formaldehyde (H\mol{2}C=O) &     -0.063 &   -0.101 \\
                                          Methanol (CH\mol{3}-OH) &     -0.096 &   -0.130 \\
                                                         N\mol{2} &     -0.030 &   -0.071 \\
                                  Hydrazine (H\mol{2}N-NH\mol{2}) &     -0.099 &   -0.134 \\
                                                               NO &     -0.025 &   -0.058 \\
                                                         O\mol{2} &     -0.027 &   -0.056 \\
                                        Hydrogen peroxide (HO-OH) &     -0.049 &   -0.061 \\
                                                         F\mol{2} &     -0.016 &   -0.020 \\
                                       Carbon dioxide (CO\mol{2}) &     -0.072 &   -0.126 \\
                                                        Na\mol{2} &     -0.077 &   -0.191 \\
                                                        Si\mol{2} &     -0.143 &   -0.585 \\
                                                         P\mol{2} &     -0.184 &   -0.694 \\
                                                         S\mol{2} &     -0.227 &   -0.788 \\
                                                        Cl\mol{2} &     -0.266 &   -0.891 \\
                                                             NaCl &     -0.180 &   -0.542 \\
                                           Silicon monoxide (SiO) &     -0.087 &   -0.311 \\
                                                               CS &     -0.118 &   -0.424 \\
                                                               SO &     -0.129 &   -0.417 \\
                                                              ClO &     -0.141 &   -0.461 \\
                                      Chlorine monofluoride (FCl) &     -0.146 &   -0.460 \\
                                                Si\mol{2}H\mol{6} &     -0.223 &   -0.661 \\
                                    Methyl chloride (CH\mol{3}Cl) &     -0.203 &   -0.542 \\
                                       Methanethiol (H\mol{3}CSH) &     -0.194 &   -0.505 \\
                                         Hypochlorous acid (HOCl) &     -0.158 &   -0.476 \\
                                       Sulfur dioxide (SO\mol{2}) &     -0.168 &   -0.469 \\
                                                        BF\mol{3} &     -0.094 &   -0.125 \\
                                                       BCl\mol{3} &     -0.453 &   -1.429 \\
                                                       AlF\mol{3} &     -0.140 &   -0.284 \\
                                                      AlCl\mol{3} &     -0.499 &   -1.606 \\
                                 Carbon tetrafluoride (CF\mol{4}) &     -0.146 &   -0.194 \\
                                Carbon tetrachloride (CCl\mol{4}) &     -0.598 &   -1.888 \\
                                       Carbon oxide sulfide (COS) &     -0.163 &   -0.484 \\
                                     Carbon bisulfide (CS\mol{2}) &     -0.256 &   -0.847 \\
                                 Carbonic difluoride (COF\mol{2}) &     -0.109 &   -0.160 \\
                               Silicon tetrafluoride (SiF\mol{4}) &     -0.200 &   -0.386 \\
                              Silicon tetrachloride (SiCl\mol{4}) &     -0.679 &   -2.169 \\
                                  Dinitrogen monoxide (N\mol{2}O) &     -0.060 &   -0.125 \\
                                   Nitrogen chloride oxide (ClNO) &     -0.162 &   -0.509 \\
                                 Nitrogen trifluoride (NF\mol{3}) &     -0.076 &   -0.116 \\
                                                        PF\mol{3} &     -0.181 &   -0.429 \\
                                                         O\mol{3} &     -0.032 &   -0.077 \\
                                                        F\mol{2}O &     -0.028 &   -0.046 \\
                                Chlorine trifluoride (ClF\mol{3}) &     -0.190 &   -0.509 \\
                         Tetrafluoro Ethene (F\mol{2}C=CF\mol{2}) &     -0.165 &   -0.244 \\
                           Acetonitrile, trifluoro- (CF\mol{3}CN) &     -0.159 &   -0.253 \\
                                       Propyne (C\mol{3}H\mol{4}) &     -0.119 &   -0.210 \\
                                        Allene (C\mol{3}H\mol{4}) &     -0.119 &   -0.208 \\
                                  Cyclopropene (C\mol{3}H\mol{4}) &     -0.117 &   -0.221 \\
                                     Propylene (C\mol{3}H\mol{6}) &     -0.154 &   -0.245 \\
                                  Cyclopropane (C\mol{3}H\mol{6}) &     -0.160 &   -0.263 \\
                                       Propane (C\mol{3}H\mol{8}) &     -0.188 &   -0.283 \\
                           Trans-1,3-butadiene (C\mol{4}H\mol{6}) &     -0.177 &   -0.298 \\
                             Dimethylacetylene (C\mol{4}H\mol{6}) &     -0.181 &   -0.305 \\
                         Methylenecyclopropane (C\mol{4}H\mol{6}) &     -0.188 &   -0.322 \\
                          Bicyclo[1.1.0]butane (C\mol{4}H\mol{6}) &     -0.190 &   -0.341 \\
                                   Cyclobutene (C\mol{4}H\mol{6}) &     -0.193 &   -0.321 \\
                                   Cyclobutane (C\mol{4}H\mol{8}) &     -0.227 &   -0.357 \\
                                     Isobutene (C\mol{4}H\mol{8}) &     -0.213 &   -0.339 \\
                                  Trans-butane(C\mol{4}H\mol{10}) &     -0.246 &   -0.375 \\
                                    Isobutane (C\mol{4}H\mol{10}) &     -0.246 &   -0.376 \\
                                  Spiropentane (C\mol{5}H\mol{8}) &     -0.258 &   -0.438 \\
                                       Benzene (C\mol{6}H\mol{6}) &     -0.263 &   -0.452 \\
                              Difluoromethane (CH\mol{2}F\mol{2}) &     -0.100 &   -0.136 \\
                                     Trifluoromethane(CHF\mol{3}) &     -0.122 &   -0.163 \\
                                               CH\mol{2}Cl\mol{2} &     -0.340 &   -0.989 \\
                                                      CHCl\mol{3} &     -0.466 &   -1.438 \\
                                Methylamine (H\mol{3}C-NH\mol{2}) &     -0.115 &   -0.163 \\
                                      Acetonitrile (CH\mol{3}-CN) &     -0.105 &   -0.186 \\
                               Nitromethane (CH\mol{3}-NO\mol{2}) &     -0.138 &   -0.225 \\
                                 Methyl nitrite (CH\mol{3}-O-N=O) &     -0.123 &   -0.200 \\
                              Methyl silane (CH\mol{3}SiH\mol{3}) &     -0.173 &   -0.423 \\
                                             Formic acid (HCOOH)  &     -0.102 &   -0.148 \\
                                   Methyl formate (HCOOCH\mol{3}) &     -0.154 &   -0.236 \\
                                 Acetamide (CH\mol{3}CONH\mol{2}) &     -0.182 &   -0.274 \\
                                                  Cyanogen (NCCN) &     -0.076 &   -0.171 \\
                             Dimethylamine ((CH\mol{3})\mol{2}NH) &     -0.173 &   -0.254 \\
                   Trans ethylamine (CH\mol{3}CH\mol{2}NH\mol{2}) &     -0.174 &   -0.255 \\
                                             Ketene (CH\mol{2}CO) &     -0.095 &   -0.165 \\
                                     Oxirane (C\mol{2}H\mol{4}O)  &     -0.125 &   -0.207 \\
                                     Acetaldehyde (CH\mol{3}CHO)  &     -0.124 &   -0.196 \\
                                                Glyoxal (HCOCOH)  &     -0.119 &   -0.198 \\
                                  Ethanol (CH\mol{3}CH\mol{2}OH)  &     -0.154 &   -0.223 \\
                             Dimethylether (CH\mol{3}OCH\mol{3})  &     -0.150 &   -0.220 \\
                                     Thiirane (C\mol{2}H\mol{4}S) &     -0.226 &   -0.575 \\
                        Dimethyl sulfoxide ((CH\mol{3})\mol{2}SO) &     -0.289 &   -0.641 \\
                                 Ethanethiol (C\mol{2}H\mol{5}SH) &     -0.256 &   -0.599 \\
                           Dimethyl sulfide (CH\mol{3}SCH\mol{3}) &     -0.253 &   -0.594 \\
                                  Vinyl fluoride (CH\mol{2}=CHF)  &     -0.111 &   -0.173 \\
                              Ethyl chloride (C\mol{2}H\mol{5}Cl) &     -0.262 &   -0.635 \\
                                  Vinyl chloride (CH\mol{2}=CHCl) &     -0.231 &   -0.600 \\
                                  Acrylonitrile (CH\mol{2}=CHCN)  &     -0.130 &   -0.239 \\
                                  Acetone (CH\mol{3}COCH\mol{3})  &     -0.186 &   -0.291 \\
                                     Acetic acid (CH\mol{3}COOH)  &     -0.164 &   -0.244 \\
                                  Acetyl fluoride (CH\mol{3}COF)  &     -0.148 &   -0.225 \\
                                  CH\mol{3}COCl (acetyl chloride) &     -0.261 &   -0.646 \\
                  CH\mol{3}CH\mol{2}CH\mol{2}Cl (propyl chloride) &     -0.326 &   -0.731 \\
                             Isopropanol (CH\mol{3})\mol{2}CHOH)  &     -0.218 &   -0.310 \\
                 Methyl ethyl ether (C\mol{2}H\mol{5}OCH\mol{3})  &     -0.213 &   -0.316 \\
                             Trimethylamine ((CH\mol{3})\mol{3}N) &     -0.234 &   -0.351 \\
                                        Furan (C\mol{4}H\mol{4}O) &     -0.199 &   -0.336 \\
                                    C\mol{4}H\mol{4}S (thiophene) &     -0.303 &   -0.709 \\
                                     Pyrrole (C\mol{4}H\mol{5}N)  &     -0.226 &   -0.375 \\
                                    Pyridine (C\mol{5}H\mol{5}N)  &     -0.253 &   -0.432 \\
                                                         H\mol{2} &     -0.011 &   -0.011 \\
                                                               HS &     -0.118 &   -0.394 \\
                                                              CCH &     -0.040 &   -0.097 \\
                          C\mol{2}H\mol{3} (\excite{2}A$^\prime$) &     -0.074 &   -0.131 \\
                               CH\mol{3}CO (\excite{2}A$^\prime$) &     -0.105 &   -0.176 \\
                                        H\mol{2}COH (\excite{2}A) &     -0.078 &   -0.110 \\
                                CH\mol{3}O (\excite{2}A$^\prime$) &     -0.076 &   -0.111 \\
               CH\mol{3}CH\mol{2}O (\excite{2}A$^{\prime\prime}$) &     -0.136 &   -0.205 \\
                                CH\mol{3}S (\excite{2}A$^\prime$) &     -0.175 &   -0.483 \\
                          C\mol{2}H\mol{5} (\excite{2}A$^\prime$) &     -0.110 &   -0.168 \\
                      (CH\mol{3})\mol{2}CH (\excite{2}A$^\prime$) &     -0.171 &   -0.263 \\
                      (CH\mol{3})\mol{2}CH (\excite{2}A$^\prime$) &     -0.171 &   -0.263 \\
                                                        NO\mol{2} &     -0.058 &   -0.114 \\
                                 Methyl allene (C\mol{4}H\mol{6}) &     -0.177 &   -0.300 \\
                                      Isoprene (C\mol{5}H\mol{8}) &     -0.238 &   -0.394 \\
                                Cyclopentane (C\mol{5}H\mol{10})  &     -0.295 &   -0.460 \\
                                    n-Pentane (C\mol{5}H\mol{12}) &     -0.310 &   -0.472 \\
                                  Neo pentane (C\mol{5}H\mol{12}) &     -0.307 &   -0.470 \\
                            1,4 Cyclohexadiene (C\mol{6}H\mol{8}) &     -0.294 &   -0.487 \\
                                  Cyclohexane (C\mol{6}H\mol{12}) &     -0.360 &   -0.561 \\
                                     n-Hexane (C\mol{6}H\mol{14}) &     -0.366 &   -0.563 \\
                             3-Methyl pentane (C\mol{6}H\mol{14}) &     -0.366 &   -0.563 \\
                                    n-Heptane (C\mol{7}H\mol{16}) &     -0.426 &   -0.657 \\
                             Cyclooctatetraene (C\mol{8}H\mol{8}) &     -0.345 &   -0.598 \\
                                     n-Octane (C\mol{8}H\mol{18}) &     -0.486 &   -0.751 \\
                                  Naphthalene (C\mol{10}H\mol{8}) &     -0.435 &   -0.755 \\
                                     Azulene (C\mol{10}H\mol{8})  &     -0.429 &   -0.747 \\
                 Acetic acid methyl ester (CH\mol{3}COOCH\mol{3}) &     -0.219 &   -0.332 \\
                               t-Butanol ((CH\mol{3})\mol{3}COH)  &     -0.276 &   -0.413 \\
                              Aniline (C\mol{6}H\mol{5}NH\mol{2}) &     -0.317 &   -0.526 \\
                                      Phenol (C\mol{6}H\mol{5}OH) &     -0.297 &   -0.494 \\
                               Divinyl ether (C\mol{4}H\mol{6}O)  &     -0.206 &   -0.338 \\
                             Tetrahydrofuran (C\mol{4}H\mol{8}O)  &     -0.264 &   -0.403 \\
                              Cyclopentanone (C\mol{5}H\mol{8}O)  &     -0.297 &   -0.473 \\
                            Pyrimidine (C\mol{4}H\mol{4}N\mol{2}) &     -0.241 &   -0.409 \\
                    Dimethyl sulphone (C\mol{2}H\mol{6}O\mol{2}S) &     -0.342 &   -0.715 \\
                               Chlorobenzene (C\mol{6}H\mol{5}Cl) &     -0.402 &   -0.902 \\
                     Butanedinitrile (NC-CH\mol{2}-CH\mol{2}-CN)  &     -0.204 &   -0.368 \\
                              Pyrazine (C\mol{4}H\mol{4}N\mol{2}) &     -0.238 &   -0.406 \\
                          Acetyl acetylene (CH\mol{3}-C(=O)-CCH)  &     -0.174 &   -0.308 \\
                            Crotonaldehyde (CH\mol{3}-CH=CH-CHO)  &     -0.211 &   -0.346 \\
             Acetic anhydride (CH\mol{3}-C(=O)-O-C(=O)-CH\mol{3}) &     -0.277 &   -0.438 \\
                         2,5-Dihydrothiophene (C\mol{4}H\mol{6}S) &     -0.330 &   -0.738 \\
                      Isobutane nitrile ((CH\mol{3})\mol{2}CH-CN) &     -0.228 &   -0.374 \\
          Methyl ethyl ketone (CH\mol{3}-CO-CH\mol{2}-CH\mol{3})  &     -0.250 &   -0.390 \\
                           Isobutanal ((CH\mol{3})\mol{2}CH-CHO)  &     -0.244 &   -0.383 \\
                          1,4-Dioxane (C\mol{4}H\mol{8}O\mol{2})  &     -0.288 &   -0.437 \\
                          Tetrahydrothiophene (C\mol{4}H\mol{8}S) &     -0.364 &   -0.775 \\
                        t-Butyl chloride ((CH\mol{3})\mol{3}C-Cl) &     -0.384 &   -0.825 \\
    n-Butyl chloride (CH\mol{3}-CH\mol{2}-CH\mol{2}-CH\mol{2}-Cl) &     -0.382 &   -0.823 \\
                          Tetrahydropyrrole (C\mol{4}H\mol{8}NH)  &     -0.286 &   -0.438 \\
    Nitro-s-butane (CH\mol{3}-CH\mol{2}-CH(CH\mol{3})-NO\mol{2})  &     -0.322 &   -0.509 \\
       Diethyl ether (CH\mol{3}-CH\mol{2}-O-CH\mol{2}-CH\mol{3})  &     -0.270 &   -0.408 \\
                Dimethyl acetal (CH\mol{3}-CH(OCH\mol{3})\mol{2}) &     -0.296 &   -0.441 \\
                           t-Butanethiol ((CH\mol{3})\mol{3}C-SH) &     -0.378 &   -0.788 \\
  Diethyl disulfide (CH\mol{3}-CH\mol{2}-S-S-CH\mol{2}-CH\mol{3}) &     -0.501 &   -1.189 \\
                    t-Butylamine ((CH\mol{3})\mol{3}C-NH\mol{2})  &     -0.296 &   -0.444 \\
                         Tetramethylsilane (Si(CH\mol{3})\mol{4}) &     -0.354 &   -0.703 \\
                           2-Methyl thiophene (C\mol{5}H\mol{6}S) &     -0.364 &   -0.804 \\
               N-methyl pyrrole (cyc-C\mol{4}H\mol{4}N-CH\mol{3}) &     -0.286 &   -0.468 \\
                             Tetrahydropyran (C\mol{5}H\mol{10}O) &     -0.324 &   -0.499 \\
     Diethyl ketone (CH\mol{3}-CH\mol{2}-CO-CH\mol{2}-CH\mol{3})  &     -0.306 &   -0.480 \\
       Isopropyl acetate (CH\mol{3}-C(=O)-O-CH(CH\mol{3})\mol{2}) &     -0.336 &   -0.519 \\
                         Tetrahydrothiopyran (C\mol{5}H\mol{10}S) &     -0.428 &   -0.875 \\
          t-Butyl methyl ether ((CH\mol{3})\mol{3}C-O-CH\mol{3})  &     -0.331 &   -0.502 \\
                  1,3-Difluorobenzene (C\mol{6}H\mol{4}F\mol{2})  &     -0.303 &   -0.496 \\
                  1,4-Difluorobenzene (C\mol{6}H\mol{4}F\mol{2})  &     -0.303 &   -0.497 \\
Di-isopropyl ether ((CH\mol{3})\mol{2}CH-O-CH(CH\mol{3})\mol{2})  &     -0.391 &   -0.595 \\
                                                        PF\mol{5} &     -0.248 &   -0.509 \\
                                                        SF\mol{6} &     -0.302 &   -0.613 \\
                                                         P\mol{4} &     -0.400 &   -1.449 \\
                                                        SO\mol{3} &     -0.213 &   -0.534 \\
                                                       SCl\mol{2} &     -0.391 &   -1.301 \\
                                                      POCl\mol{3} &     -0.577 &   -1.782 \\
                                                       PCl\mol{5} &     -0.832 &   -2.657 \\
                                               Cl\mol{2}O\mol{2}S &     -0.477 &   -1.417 \\
                                                       PCl\mol{3} &     -0.528 &   -1.723 \\
                                                Cl\mol{2}S\mol{2} &     -0.514 &   -1.712 \\
                                              SiCl\mol{2} singlet &     -0.359 &   -1.209 \\
                                                      CF\mol{3}Cl &     -0.260 &   -0.614 \\
                             Hexafluoro ethane (C\mol{2}F\mol{6}) &     -0.239 &   -0.328 \\
                                                        CF\mol{3} &     -0.102 &   -0.143 \\
                                C\mol{6}H\mol{5} (phenyl radical) &     -0.251 &   -0.439 \\
                          Bicyclo[1.1.0]butane (C\mol{4}H\mol{6}) &     -0.190 &   -0.341 \\
                            (CH\mol{3})\mol{3}C (t-butyl radical) &     -0.232 &   -0.358 \\
                                  Trans-butane(C\mol{4}H\mol{10}) &     -0.246 &   -0.375 \\
                            1,3 Cyclohexadiene (C\mol{6}H\mol{8}) &      0.000 &   -0.485 \\
                                                               ME &     -0.209 &   -0.448 \\
                                                              MAE &      0.209 &    0.448 \\
\end{longtable}

\clearpage{}
\section{G3 set results}
\begin{table}[h!]
\caption{G3 results for open shell systems from the combined and spin-scaled models with and without Lieb-Oxford bound (LO). The values are compared against the reference values. For completeness, parent SCAN values are also presented. All values are in kcal mol$^{-1}$. All SCAN calculations were performed fully self-consistently with the 6-311++G(3df,3pd) basis set in QUEST. The ML calculations are performed non self-consistently.}
\begin{center}
\begin{tabular}{lc c c c c c c}
\hline\hline
&Molecules&Benchmark & SCAN& Combined & Spin-scaled & Combined-LO&Spin-scaled-LO\\
\hline
&CH$_{3}$CO&581.38&588.82&580.16&583.38&579.58&584.33\\
&CH$_{3}$CH$_{2}$O&696.7&702.92&688.32&702.08&689.21&698.67\\
&BeH &49.91&60.31&59.91&59.62&57.28&62.22\\
&C$_{2}$H$_{3}$&444.82&450.60&450.56&446.84&450.30&445.04\\
&C$_{2}$H$_{5}$&603.36&610.21&599.40&599.73&598.46&599.25\\
&CCH&265.64&265.41&270.57&268.17&278.08&266.29\\
&CF$_{3}$&343.59&355.42&344.54&344.89&327.37&358.13\\
&CH&83.91&81.76&84.41&83.61&83.39&85.44\\
&CH$_{3}$&307.51&312.18&305&304.69&301.75&303.99\\
&CH$_{3}$O&400.38&408.23&396.44&400.11&395.39&406.65\\
&CH$_{3}$S&380.9&384.84&382.14&382.84&386.26&385.09\\
&ClO&64.53&69.75&71.75&72.59&75.00&77.59\\
&CN&181.38&177.76&183.81&187.87&190.01&190.73\\
&H$_{2}$COH&409.42&415.51&403.63&405.53&399.97&410.11\\
&HCO&278.73&283.23&277.95&278.71&274.35&282.62\\
&NH$_{2}$&181.78&184.45&172.27&182.87&173.42&185.47\\
&NO&152.68&151.65&139.57&152&144.73&163.06\\
&NO$_{2}$&227.42&243.20&228.83&240.11&234.56&253.63\\
&O$_{2}$&120.31&128.11&106.94&119.07&115.88&128.35\\
&OH&106.43&109.19&99.13&105.22&98.54&108.18\\
&PH$_{2}$&152.93&156.87&149.14&155.72&146.01&161.72\\
&Phenyl radical&1247.48&1263.75&1258.96&1257.09&1272.19&1252.49\\
&S$_{2}$&101.62&109.32&107.92&107.70&114.03&102.89\\
&SH&86.85&89.11&88.59&89&90.43&91.68\\
&SiH$_{3}$&225.33&229.64&227.75&228.97&222.97&239.12\\
&SO$_{3}$&125.02&132.10&123.57&127.34&124.33&126.22\\
&Tbutyl radical&1198.59&1208.50&1190.27&1191.75&1194.14&1191.29\\
\hline\hline
&MAE &-&6.23&5.17&3.73&7.26&6.35\\
 \end{tabular}
\end{center}
 \label{table:open}
\end{table}

\begin{table}[h]
\caption{G3 results for closed shell systems from the combined and spin-scaled models with and without Lieb-Oxford bound (LO). The values are compared against the reference values. For completeness, parent SCAN values are also presented. All values are in kcal mol$^{-1}$. All SCAN calculations were performed fully self-consistently with the 6-311++G(3df,3pd) basis set in QUEST. The ML calculations are performed non self-consistently.}
\begin{center}
\resizebox{\columnwidth}{!}{
\begin{tabular}{lc r r r r r r}
\hline\hline
&Molecules&Benchmark & SCAN& Combined & Spin-scaled& Combined-LO&Spin-scaled-LO\\
\hline
&1Propyl-chloride&985.15&991.92&986.51&984.37&993.59&988.20\\
&2Butyne&1003.96&1005.26&1007.56&1005.74&1011.95&1004.60\\
&2Methylpropanenitrile&1206.72&1206.09&1200.49&1204.58&1210.49&1208.58\\
&2Nitrobutane&1490.14&1505.27&1484.34&1489.19&1489.83&1506.26\\
&3Methyl-pentane&1890.37&1894.20&1879.61&1877.69&1886.12&1878.21\\
&11Dimethoxyethane&1486.73&1496.45&1482.41&1478.17&1476.69&1495.90\\
&12Butadiene&999.46&1003.66&1008.46&1006.45&1011.81&1005.19\\
&13Cyclohexadiene&1477.23&1482.99&1480.31&1480.23&1491.83&1478.67\\
&13Difluorobenzene&1385.3&31405.42&1400.02&1401.14&1399.77&1409.74\\
&14Benzoquinon&1423.17&31435.89&1436.34&1444.08&1433.60&1441.10\\
&14Cyclohexadiene&1477.61&31482.67&1479.34&1490.94&1479.58&1477.47\\
&14Difluorobenzene&1385.93&31404.36&1399.58&1398.93&1400.44&1409.18\\
&14Dioxane&1354.66&1366.22&1351.77&1348.60&1350.59&1365.58\\
&25Dihydrothiophene&1085.28&1099.32&1097.08&1099.26&1113.94&1103.07\\
&Acetaldehyde&676.89&679.90&677.47&674.08&674.74&680.13\\
&Acetamide&868.15&873.52&859.95&865.45&861.62&874.90\\
&Acetic acid&802.85&809.72&796.06&795.79&791.92&807.52\\
&Aceticanhydride&1359.61&1374.53&1358.04&1357&1355.99&1372.08\\
&Acetone&978.28&983.26&977.44&974.54&976.62&980.30\\
&Acetonitrile&615.62&613.90&615.19&619.69&621.16&623.09\\
&Acetyl-fluoride&705.94&712.84&703.63&701.76&695.76&711.49\\
&Acetylacetylene&968.18&969.61&975.16&972.04&977.22&973.36\\
&Acetylene&405.54&401.46&411.06&408.30&410.52&404.57\\
&Acrylonitrile&763.97&761.65&767.27&771.17&774.94&774.01\\
&AlCl$_{3}$&306.19&312.25&317&326.90&329.68&319.18\\
&AlF$_{3}$&426.64&420.43&420.13&416.53&403.71&418.51\\
&Allene&703.29&706.34&712.78&710.36&714.25&709.10\\
&Aniline&1539.93&1551.79&1544.57&1553.23&1559.93&1556.83\\
&Aziridine&719.85&722.01&719.92&721.74&722.95&728.29\\
&Azulene&2128.44&2148.04&2151.95&2154.61&2174.51&2153.36\\
&BCl$_{3}$&322.71&332.14&333.38&344.98&351.70&340.98\\
&Benzene&1367.74&1376.96&1379.55&1379.48&1390.56&1378.91\\
&Bicyclo-butane&987.05&991.11&995.38&991.78&1003.85&992.35\\
&Butadiene-C$_{2}$H&1012.45&1014.87&1020.59&1017.11&1022.27&1017.19\\
&C$_{2}$F$_{6}$&788.37&803.80&779.18&799.15&742.22&806.96\\
&C$_{2}$Cl$_{4}$&312.55&317.90&33.45&336.03&354.63&340.34\\
&CF$_{2}$CF$_{2}$&583.81&599.58&590.92&589.68&566.97&607.89\\
&CF$_{3}$Cl&431.63&442.04&430.15&432.28&417.68&450.13\\
&CF$_{3}$CN&639.45&646.09&636.92&642.82&624.98&659.85\\
&CF$_{4}$&476.1&483.71&465.46&466.17&439.80&485.71\\
\hline\hline
& &&&&table continued\\
\end{tabular}
}
\end{center}
\end{table}

\begin{table}[h]
\begin{center}
\resizebox{\columnwidth}{!}{
\begin{tabular}{lc c c c c c c}
\hline\hline
&Molecules&Benchmark & SCAN& Combined & Spin-scaled& Combined-LO&Spin-scaled-LO\\
\hline
&CH$_{2}$Cl$_{2}$&369.7&376.70&379.39&379.40&392.05&385.30\\
&CH$_{2}$F$_{2}$&436.32&440.32&435.07&431.59&420.23&441.95\\
&CH$_{3}$Cl&394.51&397.94&398.99&397.20&402.88&401.31\\
&CH$_{3}$SH&473.6&475.67&476.63&476.58&482.16&480.29\\
&CH$_{4}$&420.13&419.38&418.10&415.34&414.65&418.25\\
&CH21A1&180.77&175.50&184.08&177.97&181.86&181.59\\
&CH23B1&190.37&196.66&201.44&189.58&199.85&190.65\\
&CHCl$_{3}$&343.14&347.19&358.69&359.66&373.78&363.70\\
&CHF$_{3}$&457.71&463.11&452.52&450.64&431.91&465.25\\
&Chlorobenzene&1346.76&1367.06&1370.24&1372.37&1391.06&1374.44\\
&Cl$_{2}$&57.91&56.76&70.46&68.86&74.49&72.35\\
&Cl$_{2}$CCCl$_{2}$&473.13&477.17&494.63&497.87&523.16&502.91\\
&ClF&61.3&60.86&72.58&67.62&66.91&71.44\\
&ClF$_{3}$&125.4&138.97&149.75&143.55&130.37&152.54\\
&ClNO&191.24&205.12&210.42&214.35&217.57&227.24\\
&CO&259.21&255.24&257.60&255.68&255.81&264.17\\
&CO$_{2}$&389.26&394.92&388.93&388.11&387.06&396.85\\
&COF$_{2}$&423.05&426.51&414.59&414.81&400.14&428.58\\
&Crotonaldehyde&1124.16&1131.64&1130.91&1127.13&1131.56&1132.22\\
&CS&423.05&426.51&414.59&414.81&190.47&188.08\\
&CS$_{2}$&277.86&287.93&301.33&305.77&320.07&312.65\\
&Cyanogen&502.26&4999.70&507.91&519.86&521.28&528.06\\
&Cyclobutane&1148.61&1152.82&1144.85&1144.79&1149.27&1150.56\\
&Cyclobutene&1001.65&1003.91&1002.83&1002.89&1007.27&1007.52\\
&Cyclohexane&1764.52&1770.18&1752.99&1753.37&1764.71&1752.50\\
&Cyclooctatetraene&1778.7&1786.04&1792.37&1791.69&1806.97&1787.84\\
&Cyclopentanetwist&1463.16&1468.05&1454.52&1455.20&1464.22&1456.58\\
&Cyclopentanone&1440.29&1445.47&1431.46&1432.03&1440.10&1436.70\\
&Cyclopropane&853.09&856.04&856.41&853.54&859.95&856.50\\
&Cyclopropene&682.88&682.45&692.31&689.17&695.61&687.46\\
&Diethylether&1392.44&1398.98&1388.52&1384.35&1385.80&1394.64\\
&Diethylketone&1566.9&1574.89&1563.43&1560.80&1566.35&1565.48\\
&Diisopropylether&1986.45&1994.61&1978.22&1974.72&1979.18&1983.61\\
&Dimethylether&797.6&801.68&796.48&792.22&791.00&803.36\\
&Dimethylamine&870.11&871.17&862.55&865.39&862.85&874.53\\
&Divinylether&1103.22&1110.33&1110.24&1107.74&1109.50&1114.38\\
&Ethane&718.29&712.69&709.62&706.91&707.98&709.19\\
&Ethanol&809.99&812.20&804.49&802.30&800.78&809.85\\
&Ethyl-chloride&691.04&701.99&700.47&699.09&692.85&704.15\\
&Ethylamine&563.47&562.67&566.88&563.52&870.20&879.81\\
&F$_{2}$&38.41&35.99&48.23&40.92&35.43&46.55\\
&F$_{2}$O&92.88&97.30&110.78&102.81&96.87&114.00\\
&Flurobenzene&1377.44&1391.45&1390.23&1390.68&1395.43&1394.66\\
\hline\hline
& &&&&table continued\\
\end{tabular}
}
\end{center}
\end{table}

\begin{table}[h]
\begin{center}
\resizebox{\columnwidth}{!}{
\begin{tabular}{lc c c c c c c}
\hline\hline
&Molecules&Benchmark & SCAN& Combined & Spin-scaled& Combined-LO&Spin-scaled-LO\\
\hline
&Furan&993.81&1002.94&100.48&1001.03&1004.89&1009.38\\
&Glyoxal&633.42&638.35&636.33&632.29&632.80&643.82\\
&H$_{2}$&109.6&107.58&106.58&104.63&103.38&108.50\\
&H$_{2}$CO&374.52&374.52&374.46&370.88&370.30&377.81\\
&H$_{2}$NNH$_{2}$&438.28&434.99&425.53&436.25&427.61&447.72\\
&H$_{2}$O&232.54&229.92&225.97&226.16&221.03&230.92\\
&H$_{2}$O$_{2}$&268.69&268.25&267.49&265.10&261.70&274.28\\
&H$_{2}$S&182.4&183.18&186.07&186.06&189.81&189.51\\
&HCl&106.42&107.59&109.14&108.36&112.66&110.99\\
&HCN&312.74&307.76&312.74&317.02&316.55&319.56\\
&HCOOCH$_{3}$-methyl&784.56&793.06&785.27&781.93&779.44&794.85\\
&HCOOH-acid&500.92&505.37&494.28&493.77&488.44&505.33\\
&HF&140.97&137.78&135.20&133.91&125.94&136.68\\
&HOCl&164.41&164.48&172.04&169.0&170.52&174.47\\
&Isobutane&1302.57&1305.01&1295.99&1293.40&1298.31&1295.14\\
&Isobutene-C$_{2}$V&1158.36&1160.96&1158.81&1156.23&1160.97&1157.07\\
&Isobutyraldehyde&1267.46&1271.43&1263.07&1259.93&1264.79&1265.63\\
&Isoprene&1310.02&1313.52&1316.01&1313.01&1319.66&1312.78\\
&Isopropyl-alcohol&1107.96&1112.09&1099.45&1097.40&1097.62&1105.49\\
&Isopropylacetate&1682&1693.40&1676.68&1674.21&1675.67&1685.85\\
&Ketene&532.45&539.08&538.91&537.54&538.87&541.78\\
&Li$_{2}$&24.42&18.46&17.16&20.69&16.82&18.57\\
&Li$_{2}$F&137.61&134.10&135.42&134.15&127.98&134.21\\
&LiH&58.03&55.49&54.47&56.82&51.72&59.19\\
&Methanol&512.48&513.68&507.17&505.70&502.71&513.95\\
&Methyl-ethyl-ether&1094.62&1100.96&1091.22&1087.02&1087.28&1098.78\\
&Methylacetate&1087&1096.88&1083.40&1081.0&1079.19&1095.62\\
&Methylamine&582.14&580.93&573.34&577.35&573.31&584.04\\
&Methylene-cyclopropane&990.6&997.69&1000.62&997.65&1006.20&999.42\\
&Methylethylketone&1272.98&1280.16&1268.65&1266.28&1243.84&1271.45\\
&Methylnitrite&598.17&605.28&600.42&603.42&600.27&623.60\\
&Methylsilane&627.2&627.71&619.99&627.72&616.27&635.15\\
&N-butane&1300.8&1303.61&1294.93&1292.34&1297.04&1294.02\\
&N$_{2}$&228.46&219.49&214.97&229.85&226.68&241.37\\
&N$_{2}$O&270.55&272.55&265.71&281.34&277.06&294.87\\
&Na$_{2}$&17.02&14.53&13.95&15.11&14.79&13.77\\
&NaCl&98.01&96.60&101.65&102.01&105.44&100.51\\
&Naphthalene&2162.08&2181.06&2182.19&2185.26&2205.87&2182.18\\
&Nbutylchloride&1286.54&1292.85&1285.80&1284.52&1295.55&1288.95\\
&Neopentane&1599.43&1602.29&1589.86&1587.51&1594.35&1589.10\\
&NF$_{3}$&204.6&212.60&215.05&215.14&196.57&231.13\\
&NH&83.57&84.81&75.16&85.74&77.16&86.84\\
\hline\hline
& &&&&table continued\\
\end{tabular}
}
\end{center}
\end{table}

\begin{table}[h]
\begin{center}
\resizebox{\columnwidth}{!}{
\begin{tabular}{lc c c c c c c}
\hline\hline
&Molecules&Benchmark & SCAN& Combined & Spin-scaled&Combined-LO&Spin-scaled-LO\\
\hline
&NH$_{3}$&297.96&294.71&288.35&293.80&287.78&298.22\\
&Nheptane&2183.81&2189.79&2172.83&2170.37&2180.58&2170.93\\
&Nhexane&1889.94&1894.38&1880.05&1877.65&1885.99&1878.70\\
&Nitromethane&600.29&609.64&598.82&600.19&603.61&620.98\\
&Nmethylpyrrole&1362.05&1370.74&1361.10&1367.62&1372.72&1377.05\\
&Noctane&2478.18&2485.18&2465.12&2462.96&2475.05&2463.39\\
&Npentane&1595.27&1599.70&1586.05&1582.69&1590.69&1585.43\\
&O$_{3}$&146.81&149.25&156.20&150.97&152.69&169.10\\
&OCS&333.55&342.11&345.79&347.53&353.40&356.56\\
&Oxirane&650.45&655.82&656.48&650.66&653.03&656.99\\
&P$_{2}$&117.26&113.86&109.65&133.30&111.65&141.94\\
&P$_{4}$&286.69&295.35&288.45&322.02&285.76&341.53\\
&PCl$_{3}$&232.54&229.92&225.97&226.16&248.61&259.37\\
&PCl$_{5}$&308.67&321.29&320.68&343.91&339.47&349.16\\
&PF$_{5}$&556.86&551.93&527.09&538.79&486.45&554.98\\
&PH$_{3}$&241.89&242.42&236.47&245.49&232.43&251.96\\
&Phenol&1471.74&1484.0&1478.98&1481.18&1488.81&1485.37\\
&Piperidine&1630.47&1636.30&1614.35&1619.60&1626.03&1626.39\\
&POCl$_{3}$&358.61&365.09&355.95&374.77&369.69&380.39\\
&Propane&1006.39&1008.59&1001.43&998.25&1001.80&1001.09\\
&Propene&860.47&861.82&861.81&858.52&862.11&859.99\\
&Propyne&704.81&704.06&709.95&707.67&711.79&705.20\\
&Pyrazine&1108.13&1114.85&1108.59&1120.04&1123.31&1131.29\\
&Pyridine&1237.7&1247.18&1244.11&1249.49&1257.00&1255.03\\
&Pyrimidine&1108.39&1119.19&1111.91&1123.51&1126.82&1134.56\\
&Pyrrole&1071.67&1078.72&1070.89&1078.67&1082.09&1085.82\\
&S$_{2}$Cl$_{2}$&194.98&205.77&225.49&230.46&240.02&232.48\\
&SF$_{6}$&477.89&487.26&464.48&469.43&417.67&496.35\\
&Si$_{2}$&74.7&73.29&68.47&74.79&66.42&72.58\\
&Si$_{2}$H$_{6}$&530.28&534.18&524.34&540.31&515.90&553.25\\
&Si$_{2}$Cl$_{2}$&205.83&205.95&207.45&223.0&216.42&219.44\\
&SCl$_{2}$&128.51&131.13&146.78&149.16&155.03&152.93\\
&SiCl$_{4}$&383.38&422.45&425.19&438.78&435.89&417.73\\
&SiH$_{4}$&321.93&322.54&317.77&324.90&311.83&334.59\\
&SiH21A1&151.43&148.62&149.32&156.98&145.47&165.00\\
&SiH23B1&130.74&137.48&142.92&134.75&138.32&146.91\\
&SiO&192.29&186.53&187.55&196.42&187.11&197.55\\
&SO$_{2}$&258.32&259.82&259.84&259.30&257.99&270.42\\
\hline\hline
& &&&&table continued\\
\end{tabular}
}
\end{center}
\end{table}

\begin{table}[t]
\begin{center}
\resizebox{\columnwidth}{!}{
\begin{tabular}{lc c c c c c c}
\hline\hline
&Molecules&Benchmark & SCAN& Combined & Spin-scaled&Combined-LO&Spin-scaled-LO\\
\hline
&SO$_{2}$Cl$_{2}$&331.72&340.67&337.78&342.21&343.85&352.92\\
&SO$_{3}$D$_{3}$H&343.41&340.67&337.78&342.21&343.85&352.92\\
&Spiropentane&1283.98&1292.18&1291.17&1287.71&1301.24&1292.06\\
&Succinonitrile&1104.13&1101.82&1102.21&1113.82&1119.67&1118.86\\
&Tbutanol&1406.15&1410.45&1396.37&1394.68&1396.19&1401.19\\
&Tbutylamine&1472.57&1474.88&1458.40&1462.89&1463.78&1469.30\\
&Tbutylmethylether&1688.5&1695.36&1681.10&1677.99&1681.15&1687.10\\
&Tetrahydrofuran&1261.54&1269.21&1254.70&1253.16&1257.92&1264.14\\
&Tetrahydropyran&1560.48&1568.99&1552.78&1551.59&1558.39&1559.82\\
&Tetrahydropyrrole&1330.43&1335.32&1317.75&1323.04&1327.05&1331.56\\
&Tetrahydrothiophene&1230.78&1244.0&1234.43&1236.87&1252.39&1241.56\\
&Tetrahydrothiopyran&1527.18&1541.86&1529.10&1531.26&1549.94&1534.06\\
&Thioethanol&768.66&778.75&776.68&775.36&786.25&780.60\\
&Thiophene&962.9&978.39&980.41&983.77&1001.46&989.03\\
&Toulene&1664.38&1674.87&1674.19&1674.59&1687.30&1673.90\\
&Trimethyl-amine&1160.14&1163.68&1152.16&1153.43&1152.81&1164.40\\
&Vinly-chloride&539.13&547.18&551.61&549.99&559.09&552.41\\
&Vinyl-fluoride&571.95&576.25&576.52&573.38&569.48&577.82\\
&2Methylthiophene&1259.49&1276.44&1275.38&1278.99&1298.40&1284.18\\
&Acetyl-chloride&667.5&683.03&679.97&679.86&685.97&688.05\\
&CH$_{3}$SCH$_{3}$&767.07&777.72&776.41&775.34&786.42&780.81\\
&Diethyldisulfide&1421.69&1443.62&1437.88&1439.50&1460.26&1448.49\\
&Dimethyl-sulfone&969.9&985.24&966.77&966.49&971.86&975.62\\
&Dimethylsulfoxide&854.96&868.14&862.42&861.0&870.34&868.51\\
&PF$_{3}$&363.92&367.55&352.65&363.70&323.86&377.69\\
&SiF$_{4}$&573.13&573.50&557.42&562.12&526.91&581.70\\
&Tbutanethiol&1361.67&1373.80&1365.02&1364.10&1377.90&1368.81\\
&Tbutylchloride&1285.63&1299.24&1291.06&1289.93&1299.65&1293.84\\
&Tetramethylsilane&1545.29&1553.12&1536.55&1547.76&1539.13&1547.42\\
&Thiirane&623.64&636.15&640.63&638.24&650.28&645.43\\
\hline\hline
&MAE &-&6.52&7.80&8.55&13.24&11.14\\
\end{tabular}
}
\end{center}
\end{table}
\label{table:closed}

\clearpage
\section{Lattice constant values for SL20}
\begin{table}[h!]
\caption{Lattice constant results from the combined and spin-scaled models with and without Lieb-Oxford bound (LO). The values are compared against the reference values. For completeness, parent SCAN values are also presented. All values are in \ang. The PBE densities were used to calculate the lattice constants using FHI-aims \cite{blum2009ab}. The ML calculations are performed non self-consistently using the PBE densities.}
\begin{center}
\begin{tabular}{lc c c c c c c}
\hline\hline
&Solids&Experimental & SCAN& Combined & Spin-scaled & Combined-LO&Spin-scaled-LO\\
\hline
&Ag &4.063&4.079&4.129&4.137&4.128&4.053\\
&Al&4.019&4.005 &3.920&4.039&3.901&4.097\\
&C&3.555&3.551&3.594&3.593&3.618&3.527\\
&Ca&5.555&5.541&5.508&5.490&5.613&5.552\\
&Cu&3.595&3.567&3.521&3.612&3.561&3.504\\
&Si&5.422&5.424&5.538&5.512&5.499&5.491\\
&Ge&5.644&5.673&5.781&5.749&5.874&5.726\\
&GaAs&5.641&5.658&5.733&5.737&5.810&5.715\\
&Na&4.207&4.190&4.328&4.116&4.355&4.199\\
&SiC&4.348&4.349&4.452&4.408&4.434&4.361\\
&Li&3.451&3.460&3.504&3.456&3.564&3.552\\
&Sr&6.042&6.085&5.968&6.018&6.051&5.970\\
&Pd&3.876&3.895&3.934&3.947&3.943&3.897\\
&LiCl&5.072&5.080&5.097&5.126&5.115&5.097\\
&MgO&4.188&4.206&4.273&4.199&4.264&4.199\\
&NaCl&5.565&5.567&5.663&5.627&5.633&5.657\\
&NaF&4.57&4.583&4.689&4.604&4.639&4.614\\
&Rh&3.793&3.789&3.811&3.834&3.836&3.818\\
&LiF&3.974&3.980&4.047&3.994&4.019&3.986\\
&Ba&5.004&5.052&4.965&5.063&4.794&4.826\\
\hline\hline
&MAE &-&0.0156&0.076&0.052&0.089&0.052\\
 \end{tabular}
\end{center}
 \label{table:sl20}
\end{table}

\clearpage

\clearpage
\bibliographystyle{ieeetr}{}
\bibliography{reference, SCANML}